\documentclass[12pt]{article}
\usepackage{a4}
\usepackage{amssymb}
\usepackage{cite}
\newcommand{\be}{\begin{equation}}
\newcommand{\ee}{\end{equation}}
\newcommand{\bea}{\begin{eqnarray}}
\newcommand{\eea}{\end{eqnarray}}

\begin{document}
\thispagestyle{empty}
\def\thefootnote{\fnsymbol{footnote}}
\vspace*{1cm}
\begin{center}\Large
Some remarks on D-branes and defects  in  Liouville and Toda field theories
\end{center}\vskip 1.5em
\begin{center}
Gor Sarkissian
\footnote{\scriptsize
~Email address: \\
$~$\hspace*{2.4em} gor.sarkissian@ysu.am
}
\end{center}
\begin{center}
Department of Theoretical Physics, \ Yerevan State University,\\
Alex Manoogian 1, 0025\, Yerevan\\
Armenia
\end{center}
\vskip 1.5em
\begin{center} July 2011 \end{center}
\vskip 2em
\begin{abstract} \noindent
In this paper we analyze the Cardy-Lewellen equation in general diagonal model.
We show that in these models it takes simple form due to some general properties of conformal field theories,
like pentagon equations and OPE associativity. This implies, that the Cardy-Lewellen
equation has simple form  also in non-rational diagonal models.
We specialize our finding to the Liouville and Toda field theories. In particular we prove,
that conjectured recently defects in Toda field theory indeed satisfy the cluster equation.
We also derive  the Cardy-Lewellen
equation in all $sl(n)$ Toda field theories and prove
that the form of  boundary states found recently in $sl(3)$ Toda field theory holds in all $sl(n)$  theories as well.

\end{abstract}

\newpage

\section{Introduction}
In the last years defects  in the Liouville and Toda field theories
have attracted some attention
\cite{Alday:2009fs,Drukker:2009id,Drukker:2010jp,Passerini:2010pr,Petkova:2009pe,Gomis:2010kv}
due to
 their  important role   as counterpart of the Wilson lines in the AGT
correspondence \cite{Alday:2009aq}.
Defects in the Liouville field theory were constructed in \cite{Sarkissian:2009aa}.
In \cite{Drukker:2010jp}, the defects in Toda field theories have been written down
generalizing the formulas for them derived in \cite{Sarkissian:2009aa}.
It  was observed in these papers that in spite of non-rational character
of these theories defects have remarkably simple form, resembling the
corresponding formulas in rational conformal field theory.
Recently also boundary states were analyzed in the $sl(3)$ Toda field theory \cite{Fateev:2010za},
and it was found that they closely related to defects found in \cite{Drukker:2010jp}.
These results hint that the simplicity of defects and branes in the Liouville and Toda field theories
dictated  by some general properties
of conformal field theory not related to rationality.
In this paper we analyze  general conditions  causing the simplicity of the Cardy-Lewellen
equation.
We show that in  diagonal theories
the pentagon equation for the fusing matrix and associativity of the operator product expansion
lead to the remarkably simple relation ( Eq. (\ref{cpijnm}) in section 1)  between the structure constant and the fusing matrix,
in turn bringing to  simple form of the Cardy-Lewellen equation.
In diagonal rational conformal field theory the mentioned relation
between the structure constant and the fusing matrix is well-known,
(see for example \cite{Behrend:1999bn,Felder:1989hq,Felder:1989wv,Fuchs:2004xi,Moore:1988ss,Moore:1989vd,Runkel:1998pm}),  but here we rederive it in a way, which does not use rationality.  Therefore
this relation should hold also in non-rational diagonal models.
 Related discussion can be found also in \cite{Petkova:2009pe}.
The paper is organized as follow.
In section 2 we derive relation between the structure constant and the fusing matrix,
taking special care on normalization of fields.
Using this relation we derive the Cardy-Lewellen equation, and show how
having a solution one can construct boundary states, permutation branes and defects.
In section 3 we consider the Liouville field theory and show how it fits to the general scheme
developed in section 2. In section 4 we consider $sl(n)$ Toda field theory, and using formalism
of section 2, derive the Cardy-Lewellen equation, describe its solutions, and present
boundary states, permutation branes and defects.

\newpage
\section{Cardy-Lewellen equations in diagonal models}
In this section we derive the relation between the structure constant and the fusing matrix in diagonal models,
which will enable us to compute the classifying algebra and write down the Cardy-Lewellen equation.
This relation is well  known in diagonal  RCFT,  where the classifying algebra structure constants
are given by the fusion coefficients \cite{Behrend:1999bn,Runkel:1998pm}.
 Here we rederive this relation in a way, which makes clear, that it is dictated
by the pentagon equation  and the OPE associativity and does not depend on rationality.
Therefore this relation in
 some way should hold also in non-rational diagonal theories.
It  explains why even in non-rational theories,
discussed in last years, like  Liouville and  Toda field theories,
simple formulae for defects and boundary states have been derived.

Let us collect the standard stuff on  2d CFT.  Denote by  $R_i$  the highest weight representations. Denote by ${\cal T}$  the set of all  $R_i$ of the CFT in question. In this paper we consider non-rational 2d CFT, i.e. we allow
the set ${\cal T}$ to be infinite. Writing $\sum_i $ we understand  the sum over all the set ${\cal T}$.
As usual, in the case of the continuous set  ${\cal T}$ the sum should be understood as an integral, the Kronecker delta as the Dirac delta function etc.
 $N_{ij}^k$ are fusion coefficients.   The vacuum representation is indexed by $i=0$, and $i^*$ refers to the conjugate
representation in a sense $N_{ii^*}^0$=1.

It is convenient  to introduce structure constants $C_{(i\bar{i})(j\bar{j})a\bar{a}}^{(k\bar{k})}$ via
full plane chiral decomposition of the physical fields \cite{Moore:1988qv,Moore:1989vd}:
\be\label{decompm}
\Phi_{(i\bar{i})}(z,\bar{z})=\sum_{j,\bar{j},k,\bar{k},a,\bar{a}}C_{(i\bar{i})(j\bar{j})a\bar{a}}^{(k\bar{k})}\left(\phi_{ija}^k(z)\otimes \phi_{\bar{i}\bar{j}\bar{a}}^{\bar{k}}(\bar{z})\right)\, ,
\ee
where $\phi_{ija}^k$ are intertwining operators $R_j\rightarrow R_k$,  and $a=1\ldots N_{ij}^k$.
It is important to note that in the case of the models with multiplicities structure constants carry additional indices
$a$ and $\bar{a}$ to disentangle different channels of the fusion.
 
Bulk OPE has the form \cite{Behrend:1999bn}
\be\label{bope}
\Phi_{(i\bar{i})}(z_1,\bar{z}_1)\Phi_{(j\bar{j})}(z_2,\bar{z}_2)=\sum_{k,\bar{k}, a, \bar{a}}
{C_{(i\bar{i})(j\bar{j})a\bar{a}}^{(k\bar{k})} \over (z_1-z_2)^{\Delta_i+\Delta_j-\Delta_k}
(\bar{z}_1-\bar{z}_2)^{\Delta_{\bar{i}}+\Delta_{\bar{j}}-\Delta_{\bar{k}}}}
\Phi_{(k\bar{k})}(z_2,\bar{z}_2)+\ldots\, .
\ee
By the usual arguments \cite{Belavin:1984vu} we have for 4-point correlation function
$\langle\Phi_i\Phi_k\Phi_j\Phi_l\rangle$
in $s$ channel
\be
\sum_{p\bar{p}}\sum_{\rho\tau\bar{\rho}\bar{\tau}}C^{p\bar{p}}_{j\bar{j}l\bar{l}(\tau\bar{\tau})}C^{i\bar{i}}_{k\bar{k}p\bar{p}(\rho\bar{\rho})}{\cal F}^s_{p\rho\tau}\left[\begin{array}{cc}
k&j\\
i&l \end{array}\right]{\cal F}^s_{\bar{p}\bar{\rho}\bar{\tau}}\left[\begin{array}{cc}
\bar{k}&\bar{j}\\
\bar{i}&\bar{l} \end{array}\right]
\ee
and $t$ channel
\be
\sum_{q\bar{q}}\sum_{\mu\nu\bar{\mu}\bar{\nu}}C^{q\bar{q}}_{k\bar{k}j\bar{j}(\mu\bar{\mu})}C^{i\bar{i}}_{q\bar{q}l\bar{l}(\nu\bar{\nu})}{\cal F}^t_{q\nu\mu}\left[\begin{array}{cc}
k&j\\
i&l \end{array}\right]{\cal F}^t_{\bar{q}\bar{\nu}\bar{\mu}}\left[\begin{array}{cc}
\bar{k}&\bar{j}\\
\bar{i}&\bar{l} \end{array}\right]\, ,
\ee
where ${\cal F}^s_{p\rho\tau}\left[\begin{array}{cc}
k&j\\
i&l \end{array}\right]$ and ${\cal F}^t_{q\nu\mu}\left[\begin{array}{cc}
k&j\\
i&l \end{array}\right]$ are $s$ and $t$  channels conformal blocks correspondingly.
Conformal blocks as well carry additional indices $\rho=1\ldots N_{kp}^i$, $\tau=1\ldots N_{jl}^p$, $\mu=1\ldots N_{kj}^q$
$\nu=1\ldots N_{ql}^i$, and similar for the right barred indices, to disentangle different fusion channels.
Conformal blocks in $s$ and $t$ channels are related by the fusing matrix
\be\label{trprb}
{\cal F}^s_{p\rho\tau}\left[\begin{array}{cc}
k&j\\
i&l \end{array}\right]=\sum_q\sum_{\nu\mu}F_{p,
q}\left[\begin{array}{cc}
k&j\\
i&l \end{array}\right]_{\rho\tau}^{\nu\mu}{\cal F}^t_{q\nu\mu}\left[\begin{array}{cc}
k&j\\
i&l \end{array}\right]\, ,
\ee
and hence one has:
\bea\label{cppqqf}
&&\sum_{p\bar{p}}\sum_{\rho\tau\bar{\rho}\bar{\tau}}C^{p\bar{p}}_{j\bar{j}l\bar{l}(\tau\bar{\tau})}C^{i\bar{i}}_{k\bar{k}p\bar{p}(\rho\bar{\rho})}F_{p,
q}\left[\begin{array}{cc}
k&j\\
i&l \end{array}\right]_{\rho\tau}^{\nu\mu}F_{\bar{p},
\bar{q}}\left[\begin{array}{cc}
\bar{k}&\bar{j}\\
\bar{i}&\bar{l} \end{array}\right]_{\bar{\rho}\bar{\tau}}^{\bar{\nu}\bar{\mu}}=\\ \nonumber
&&C^{q\bar{q}}_{k\bar{k}j\bar{j}(\mu\bar{\mu})}C^{i\bar{i}}_{q\bar{q}l\bar{l}(\nu\bar{\nu})}\, .
\eea
Using the relation \cite{Behrend:1999bn}
\be
\sum_{\bar{q},\bar{\nu},\bar{\mu}}F_{\bar{p},
\bar{q}^*}\left[\begin{array}{cc}
\bar{k}&\bar{j}\\
\bar{i}&\bar{l} \end{array}\right]_{\bar{\rho}\bar{\tau}}^{\bar{\nu}\bar{\mu}}F_{\bar{q},
s}\left[\begin{array}{cc}
\bar{j}&\bar{l}\\
\bar{k}^*&\bar{i}^* \end{array}\right]^{\gamma_1\gamma_2}_  {\bar{\mu}\bar{\nu}}=
\delta_{\bar{p}s}\delta_{\bar{\rho}\gamma_1}\delta_{\bar{\tau}\gamma_2}\, ,
\ee
Eq. (\ref{cppqqf}) can be written in the form:
\bea\label{cpfdig}
&&\sum_{p}\sum_{\rho\tau}C^{p\bar{p}}_{j\bar{j}l\bar{l}(\tau\bar{\tau})}C^{i\bar{i}}_{k\bar{k}p\bar{p}(\rho\bar{\rho})}F_{p,
q}\left[\begin{array}{cc}
k&j\\
i&l \end{array}\right]_{\rho\tau}^{\nu\mu}=\\ \nonumber
&&\sum_{\bar{q},\bar{\mu},\bar{\nu}}C^{q\bar{q}}_{k\bar{k}j\bar{j}(\mu\bar{\mu})}C^{i\bar{i}}_{q\bar{q}l\bar{l}(\nu\bar{\nu})}
F_{\bar{q}^*,
\bar{p}}\left[\begin{array}{cc}
\bar{j}&\bar{l}\\
\bar{k}^*&\bar{i}^* \end{array}\right]^{\bar{\rho}\bar{\tau}}_{\bar{\mu}\bar{\nu}}\, .
\eea
Putting in (\ref{cppqqf}) $i=\bar{i}=0$  we obtain the following useful relation:
\be\label{copc}
C_{j\bar{j}l\bar{l}(\tau\bar{\tau})}^{k^*,\bar{k}^*}C_{k\bar{k},k^*\bar{k}^*}^0=
C_{k\bar{k}j\bar{j}(\tau\bar{\tau})}^{l^*\bar{l}^*}C_{l^*\bar{l}^*,l\bar{l}}^0\, .
\ee

For  diagonal model
\be
C^{p\bar{p}}_{k\bar{k}i\bar{i}(\rho\bar{\rho})}=C^{p}_{ki(\rho\bar{\rho})}\delta_{\bar{p}p^*}\delta_{\bar{k}k^*}\delta_{\bar{i}i^*}
\ee 

Eq. (\ref{cpfdig}) takes the form:

\be\label{asseqmul2}
\sum_{\rho\tau}C^{i}_{kp(\rho\bar{\rho})}C^p_{jl(\tau\bar{\tau})}F_{p,q}\left[\begin{array}{cc}
k&j\\
i&l \end{array}\right]_{\rho\tau}^{\nu\mu}
=\sum_{\bar{\mu}\bar{\nu}}C^q_{kj(\mu\bar{\mu})}C^{i}_{ql(\nu\bar{\nu})}F_{q,p}\left[\begin{array}{cc}
k^*&i\\
j&l^* \end{array}\right]^{\bar{\tau}\bar{\rho}}_{\bar{\mu}\bar{\nu}}\, .
\ee

To derive (\ref{asseqmul2})  we also used the symmetry properties (\ref{symmmul}), reviewed in appendix A.

It is shown in appendix A that the pentagon equation for fusing matrix  \cite{Moore:1988qv,Moore:1989vd,Behrend:1999bn}
\bea\label{pent1ab}
\sum_{s,\beta_2, t_2,t_3}F_{p_2,s}\left[\begin{array}{cc}
j&k\\
p_1&b \end{array}\right]^{\beta_2t_3}_{\alpha_2\alpha_3}F_{p_1,l}\left[\begin{array}{cc}
i&s\\
a&b \end{array}\right]^{\gamma_1t_2}_{\alpha_1\beta_2}F_{s,r}\left[\begin{array}{cc}
i&j\\
l&k \end{array}\right]^{u_2u_3}_{t_2t_3}=\\ \nonumber
\sum_{\beta_1}
F_{p_1,r}\left[\begin{array}{cc}
i&j\\
a&p_2 \end{array}\right]^{\beta_1u_3}_{\alpha_1\alpha_2}F_{p_2,l}\left[\begin{array}{cc}
r&k\\
a&b \end{array}\right]^{\gamma_1u_2}_{\beta_1\alpha_3}
\eea
  implies  the following important relation:
\bea\label{pent6}
\sum_{\rho,\tau}
F_{0,i}\left[\begin{array}{cc}
p&k\\
p&k^* \end{array}\right]^{ \bar{\rho}\rho}_{00}
F_{p,q}\left[\begin{array}{cc}
k&j\\
i&l \end{array}\right]^{\nu\mu}_{\rho\tau}
F_{0,p}\left[\begin{array}{cc}
l&j\\
l&j^* \end{array}\right]^{\bar{\tau}\tau}_{00}
=\\ \nonumber
\sum_{\bar{\mu},\bar{\nu}}
F_{0,q}\left[\begin{array}{cc}
j&k\\
j&k^* \end{array}\right]^{\bar{\mu}\mu}_{00}
F_{q,p}\left[\begin{array}{cc}
k^*&i\\
j&l^* \end{array}\right]^{\bar{\tau} \bar{\rho}}_{\bar{\mu}\bar{\nu}}
F_{0,i}\left[\begin{array}{cc}
q&l\\
q&l^* \end{array}\right]^{\bar{\nu}\nu}_{00}\, .
\eea
It is important to note that  all the steps performed in appendix A to derive (\ref{pent6})  from (\ref{pent1ab}),
are valid as in rational as well in non-rational theories, namely all manipulations work also for infinite set ${\cal T}$ and
infinite fusion coefficients $N^i_{jk}$.
 Therefore the relation  (\ref{pent6})  holds  in non-rational theories as well.



Comparing  (\ref{pent6})  and  (\ref{asseqmul2})  we  see that (\ref{asseqmul2}) can be solved by an ansatz
\be\label{cpijnm}
C^p_{ij(\mu\bar{\mu})}={\eta_i\eta_j\over \eta_0\eta_p} F_{0,p}\left[\begin{array}{cc}
j&i\\
j&i^* \end{array}\right]^{\bar{\mu}\mu}_{00}
\ee
with arbitrary $\eta_i$.
To find $\eta_i$ we set $p=0$ 
\be\label{p0}
C^0_{ii^*}={\eta_i\eta_{i^*}\over \eta_0^2}F_i\, ,
\ee
where
\be
F_i\equiv F_{0,0}\left[\begin{array}{cc}
i&i^*\\
i&i \end{array}\right]\, .
\ee
Using 
\be
C^0_{ii^*}={C_{ii^*}\over C_{00}}\, ,
\ee
 where $C_{ii^*}$ are two-point functions and that $F_0=1$ one can solve (\ref{p0}) setting
\be\label{etai}
\eta_i=\epsilon_i\sqrt{C_{ii^*}/F_i}\, ,
\ee
were  $\epsilon_i$  a sign factor. We assume that $\epsilon_{i}$ can be chosen to satisfy $\epsilon_{i}=\epsilon_{i^*}$.

For diagonal models without multiplicities we can derive the relation (\ref{cpijnm}) in the different way.
For these models the associativity  condition (\ref{asseqmul2}) takes the form 

\be\label{asseq}
C^{p^*}_{ki^*}C^p_{jl}C^0_{pp^*}F_{p,q}\left[\begin{array}{cc}
k&j\\
i&l \end{array}\right]=C^q_{kj}C^{q^*}_{i^*l}C^0_{qq^*}F_{q,p}\left[\begin{array}{cc}
k^*&i\\
j&l^* \end{array}\right]\, .
\ee
To derive (\ref{asseq}) we used (\ref{copc}) and the commutativity of the structure constants by two lower indices 
in diagonal models  \cite{Behrend:1999bn}:
\be
C_{ik,c\bar{c}}^{j}=
C_{ki,c\bar{c}}^{j}\, .
\ee

Setting $q=0$, $k=j^*$, $i=l$ in (\ref{asseq})
we obtain:
\be\label{cfof}
\left(C^p_{ij}\right)^2={C_{jj^*}C_{ii^*}F_{0,p}\left[\begin{array}{cc}
j&i\\
j&i^* \end{array}\right]\over C_{00}C_{pp^*}F_{p,0}\left[\begin{array}{cc}
j^*&j\\
i&i \end{array}\right]}\, .
\ee

Using the relation
\be\label{foifio}
F_{0,i}\left[\begin{array}{cc}
j&k\\
j&k^* \end{array}\right]F_{i,0}\left[\begin{array}{cc}
k^*&k\\
j&j \end{array}\right]={F_jF_k\over F_i}\, ,
\ee
obtained in appendix A again as a consequence of the pentagon equation, we can write  (\ref{cfof}) in two forms
\be\label{cpijnx}
C^p_{ij}={\eta_i\eta_j\over \eta_0\eta_p} F_{0,p}\left[\begin{array}{cc}
j&i\\
j&i^* \end{array}\right]\, ,
\ee
and
\be\label{cpij}
C^p_{ij}={\xi_i\xi_j\over \xi_0\xi_p}{1\over F_{p,0}\left[\begin{array}{cc}
j^*&j\\
i&i \end{array}\right]}\, ,
\ee
where  $\eta_i$ is defined in (\ref{etai}) and 
\be
\xi_i=\eta_iF_i=\epsilon_i\sqrt{C_{ii^*}F_i}\, .
\ee
Eq. (\ref{cfof}) determines (\ref{cpijnx}) and (\ref{cpij}) only up to sign, but comparison  with (\ref{cpijnm}) shows that the sign ambiguity can be absorbed in factors $\epsilon_i$.

The relation (\ref{cpijnm})  enables as to solve the Cardy-Lewellen cluster equations for various D-branes and defects.
The Cardy-Lewellen cluster condition
for one-point functions in the presence of boundary
\be
\langle\Phi_{(i\bar{i})}(z,\bar{z})\rangle={U^i\delta_{i^*\bar{i}}\over |z-\bar{z}|^{2\Delta_i}}
\ee

 reads \cite{Behrend:1999bn}
\be\label{clmul}
\sum_{k, a, \bar{a}} C_{(ii^*)(jj^*)a\bar{a}}^{(k,k^*)}U^kF_{k0}\left[\begin{array}{cc}
i^*& i\\
j& j\end{array}\right]_{\bar{a}a}^{00}=U^iU^j\, .
\ee

Putting  (\ref{cpijnm}) in (\ref{clmul}), and using formulas (\ref{fio}) and (\ref{pent3a}) in appendix A  to perform the  sums by $a$ and $\bar{a}$,
we obtain

\be\label{nbkt}
\sum_k U^k N_{ij}^k{\xi_i\xi_j\over \xi_0\xi_k}=U^iU^j\,,
\ee
where $N_{ij}^k$ are  the fusion coefficients.
Defining
\be
U^k=\Psi^k{\xi_k\over \xi_0}
\ee
one can write (\ref{nbkt}) in the form:
\be\label{nbk}
\sum_k \Psi^k N_{ij}^k=\Psi^i\Psi^j\, .
\ee


It was shown in \cite{Sarkissian:2009aa} that the cluster condition  for two-point functions in the presence of permutation branes on two-fold product of  diagonal models
\be
\langle\Phi_{(ii^*)}^{(1)}(z_1)\Phi_{(jj^*)}^{(2)}(z_2)\rangle_{\cal P}={U^{i}_{(2){\cal P}}
\delta_{ij}
\over |z_1-\bar{z}_2|^{2\Delta_i}|\bar{z}_1-z_2|^{2\Delta_{i}}}
\ee

 is:

\bea\label{clper}
\sum_{k,,a, \bar{a},c,\bar{c}} C_{(ii^*)(jj^*)a\bar{a}}^{(k,k^*)}
C_{(ii^*)(jj^*)c\bar{c}}^{(k,k^*)}F_{k0}
 \left[\begin{array}{cc}
i^*&i\\
j&j\end{array}\right]_{\bar{c}a}^{00}F_{k0}
 \left[\begin{array}{cc}
i^*&i\\
j&j\end{array}\right]_{\bar{a}c}^{00}
U^{k}_{(2){\cal P}}=\\ \nonumber
U^{i}_{(2){\cal P}}U^{j}_{(2){\cal P}}\, .
\eea

Performing the same steps we obtain:

\be\label{2bktp}
\sum_k U^k_{(2){\cal P}} N_{ij}^k\left({\xi_i\xi_j\over \xi_0\xi_k}\right)^2=U^i_{(2){\cal P}}U^j_{(2){\cal P}}\,.
\ee
Eq. (\ref{2bktp})  can be solved by
the relation
\be
U^k_{(2){\cal P}}=\Psi^k\left({\xi_k\over \xi_0}\right)^2\, ,
\ee
with $\Psi^k$  satisfying (\ref{nbk}).

It can be shown that for permutation branes on the $N$-fold product, permuted by a cycle 
$(1\ldots N)$, the corresponding equation has the form:
\be\label{nbktp}
\sum_k U^k_{(N){\cal P}} N_{ij}^k\left({\xi_i\xi_j\over \xi_0\xi_k}\right)^N=U^i_{(N){\cal P}}U^j_{(N){\cal P}}\, ,
\ee
and therefore can be solved by
the relation
\be
U^k_{(N){\cal P}}=\Psi^k\left({\xi_k\over \xi_0}\right)^N\, ,
\ee
with $\Psi^k$ again satisfying (\ref{nbk}).

 In non-rational theories one should take care that
$N_{ij}^k$ are finite. Usually in non-rational theories this equation used, when one of the fields, say $j$, is degenerate,
and this condition is satisfied.

It was shown also in \cite{Sarkissian:2009aa} that two-point functions in the presence
of defect $D^k$ 
\be
\langle\Phi_{ii^*}(z_1,\bar{z}_1)X\Phi_{i^*i}(z_2,\bar{z}_2)\rangle={D^{i}\over 
(z_1-z_2)^{2\Delta_i}(\bar{z}_1-\bar{z}_2)^{2\Delta_{i}}}
\ee
satisfy folded version of the cluster condition for the permutation branes on
two-fold product and therefore given by the $U^k_{(2){\cal P}}$  divided by
the OPE coefficients $C^0_{kk^*}$:
\be
D^k=\Psi^k\left({\xi_k\over \xi_0}\right)^2{C_{00}\over C_{kk^*}}=\Psi^kF_k\, .
\ee

In  rational conformal field theory one has also the relation
\be\label{fas}
F_k={S_{00}\over S_{0k}}\, ,
\ee
where $S_{ab}$ is the matrix of the modular transformations.

In RCFT two-points functions can be normalized to 1. Therefore in RCFT $\xi_k={\sqrt{S_{00}}\over \sqrt{S_{0k}}}$.
 Eq. (\ref{nbk}) is solved by
\be
\Psi^k_a={S_{ak}\over S_{0a}}\, .
\ee
Taking into account the relation between one-point functions $U^k$
and coefficients of the boundary state $B^k$
\be\label{ub}
U^k={B^k\over B^0}\, ,
\ee

we obtain the formulae for the Cardy states \cite{Cardy:1989ir}:
\be
B^k_a={S_{ak}\over \sqrt{S_{0k}}}\, ,
\ee
\be
|a\rangle=\sum_k B^k_{\alpha}|k\rangle\rangle\, ,
\ee
  where $|k\rangle\rangle=\sum_N |k,N\rangle\otimes U\overline{|k,N\rangle}$  are Ishibashi states,

permutation branes \cite{Recknagel:2002qq}:
\be
B^{(N)k}_{{\cal P}a}={S_{ak}\over (S_{0k})^{N/2}}\, ,
\ee
\begin{equation}
\label{carst}
|a\rangle_{{\cal P}}=\sum_{k}{S_{ak}\over (S_{0k})^{N/2}}|k,k\rangle\rangle_{{\cal P}}\, ,
\end{equation}
where $|k,k\rangle\rangle_{{\cal P}}$ are permuted Ishibashi states \cite{Recknagel:2002qq},

and defects \cite{Petkova:2000ip} :
\be
{\cal D}^k_a={S_{ak}\over S_{0k}}\, ,
\ee
\be\label{xpd}
X=\sum_{k}{\cal D}^{k}P^{k}\, ,
\ee
where
\be
P^{k}=\sum_{N,\bar{N}}(|k,N\rangle\otimes |k^*,\bar{N}\rangle)
(\langle k,N|\otimes \langle k^*,\bar{N}|)
\ee
correspondingly.  We denoted by $|k,N\rangle$ the orthogonal basis of the highest weight representation $k$ and $U$ is an antiunitary operator
acting on $k$ by conjugation.

One can hope that (\ref{fas})  holds in non-rational theories as well,
since it reflects the equality of two expressions for the quantum dimension computed in two different ways
\cite{Moore:1988qv,Moore:1989vd}.

Now let us explain how a continuous family of the boundary states arises in non-rational conformal field theory
 \cite{Fateev:2000ik,Schomerus:2005aq}. Recall that the Cardy-Lewellen equation (\ref{clmul}) is derived by considering the
two point function $\langle \Phi_{i}(z_1) \Phi_{j}(z_2)\rangle$ in the presence of a boundary in two pictures.
In the  first picture the limit $z_1\rightarrow z_2$ is considered, using the bulk OPE (\ref{bope}).  It brings to the l.h.s. in (\ref{clmul}) (after applying the conformal block transformation rule (\ref{trprb})).
In the second picture the limit $z_1\rightarrow \bar{z}_1$ and $z_2\rightarrow \bar{z}_2$ is considered using the bulk-boundary OPE
\be\label{bboo}
\Phi_{i}(z,\bar{z})\sim \sum_{k}R(i,k)\Psi_k(x)\, ,
\ee
where $\Psi_k$ are boundary fields. Identifying $U_i=R(i,0)$ one obtains for the r.h.s. $U_iR(j,0) $.
To obtain boundary states in non-rational CFT one can use the ``Teschner trick".
Assume that $j$ is a degenerate primary producing finite number of the primary fields in the fusion with a generic primary $i$.
In this case both OPE's  (\ref{bope}) and (\ref{bboo}) contain  finite number of the primary fields in the r.h.s.
and one again reaches to the l.h.s. of the Cardy-Lewellen equation (\ref{nbkt}) with the finite sum over $k$ and the finite fusion coefficients $N_{ij}^k$. But in non-rational CFT  we have two ways to continue. One can again set
$R(j,0)=U_j$ and receive the Cardy-Lewellen equation in the form (\ref{nbkt}).
Alternatively one can treat   $R(j,0)$ (or if it diverges its leading singularity)
as a constant parameter
 $A$  characterizing a boundary condition. Setting $R(j,0)=A$ one gets linear equation
\be\label{nbkc}
\sum_k \Lambda^k N_{ij}^k=\Lambda^i A{\xi_0\over \xi_j}\, ,
\ee
where
\be
U^k=\Lambda^k\xi_k\, .
\ee
Solutions of the eq. (\ref{nbkc}) bring to the continuous family of boundary  states.

Correspondingly the continuous family of the  $N$-fold permutation branes is given by solution of the equation
\be\label{nbkc2}
\sum_k \Lambda_{(N){\cal P}}^k N_{ij}^k=\Lambda_{(N){\cal P}}^i A\left({\xi_0\over \xi_j}\right)^N\, ,
\ee
where
\be
U^k_{(N){\cal P}}=\Lambda_{(N){\cal P}}^k\xi_k^N\, ,
\ee

and the continuous family of defects, after folding of the two-fold permutation branes,  is given by the following functions
\be\label{nbkc3}
D^k=\Lambda_{(2){\cal P}}^kF_kC_{00}\, .
\ee

\section{Liouville field theory}

Let us review basic facts on the Liouville field theory (see e.g. \cite{Teschner:2001rv}).
Liouville field theory is defined on a two-dimensional surface with metric $g_{ab}$ by the local Lagrangian
density
\be
{\cal L}={1\over 4\pi}g_{ab}\partial_a\varphi\partial_b \varphi+\mu e^{2b\varphi}+{Q\over 4\pi}R\varphi\, ,
\ee
where $R$ is associated curvature. This theory is conformal invariant if the coupling constant $b$
is related with the background charge $Q$ as
\be
Q=b+{1\over b}\, .
\ee

The symmetry algebra of this conformal field theory is the Virasoro algebra
\be
[L_m,L_n]=(m-n)L_{m+n}+{c_L\over 12}(n^3-n)\delta_{n,-m}
\ee
with the central charge
\be
c_L=1+6Q^2\, .
\ee

Primary fields $V_{\alpha}$ in this theory, which are associated with exponential fields
$e^{2\alpha \varphi}$, have conformal dimensions
\be
\Delta_{\alpha}=\alpha(Q-\alpha)\, .
\ee

The spectrum of the Liouville theory is believed \cite{Curtright:1982gt,Braaten:1982yn,Braaten:1983np}
to be of the following form
\be\label{lsdi}
{\cal H}=\int_0^{\infty} dP \;R_{{Q\over 2}+iP}\otimes R_{{Q\over 2}+iP}\, ,
\ee
where $R_{\alpha}$ is the highest weight representation with respect to Virasoro algebra.
Characters of the representations $R_{{Q\over 2}+iP}$ are
\be\label{charl}
\chi_{P}(\tau)={q^{P^2}\over \eta(\tau)}\, ,
\ee
where
\be
\eta(\tau)=q^{1/24}\prod_{n=1}^{\infty}(1-q^n)\, .
\ee
Modular transformation of (\ref{charl}) is well-known:
\be\label{motrp}
\chi_{P}(-{1\over \tau})=\sqrt{2}\int\chi_{P'}(\tau)e^{4i\pi PP'}dP'\, .
\ee
Degenerate representations appear at 
\be\label{deger}
\alpha_{m,n}={1-m\over 2b}+{1-n\over 2}b
\ee
 and have
conformal dimensions
\be
\Delta_{m,n}=Q^2/4-(m/b+nb)^2/4\, ,
\ee
where $m,n$ are positive integers. At general $b$ there is only one null-vector at the level $mn$.
Hence the degenerate character reads:
\be\label{charmn}
\chi_{m,n}(\tau)={q^{-(m/b+nb)^2}-q^{-(m/b-nb)^2}\over \eta(\tau)}\, .
\ee

Modular transformation of (\ref{charmn}) is worked out in \cite{Zamolodchikov:2001ah}
\be\label{motrdi}
\chi_{m,n}(-{1\over \tau})=2\sqrt{2}\int\chi_{P}(\tau)\sinh(2\pi mP/b)\sinh(2\pi nbP)dP\, .
\ee
Given that the identity field is specified by $(m,n)=(1,1)$  one finds the vacuum component of the matrix of modular transformation:
\be
S_{0\alpha}=-i2\sqrt{2}\sin\pi b^{-1}(2\alpha-Q)\sin\pi b(2\alpha-Q)\, .
\ee

To present formula (\ref{cpij}) in the Liouville field theory we need two-point function
\be\label{twopf}
\langle V_{\alpha}(z_1,\bar{z}_1)V_{\alpha}(z_2,\bar{z}_2)\rangle={S(\alpha)\over (z_1-z_2)^{2\Delta_{\alpha}}
(\bar{z}_1-\bar{z}_2)^{2\Delta_{\alpha}}}\, .
\ee

Let us for this purpose recall some facts on the values of the correlation functions in the Liouville field theory in the Coulomb gas approach.
\begin{enumerate}
\item
The three-point functions satisfying the relation $\alpha_1+\alpha_2+\alpha_3=Q$ are set to 1.
This rule actually sets normalization  of the fields, since from here we receive that
\be\label{twopfq}
\langle V_{\alpha}(z_1,\bar{z}_1)V_{Q-\alpha}(z_2,\bar{z}_2)\rangle={1\over (z_1-z_2)^{2\Delta_{\alpha}}
(\bar{z}_1-\bar{z}_2)^{2\Delta_{\alpha}}}\, .
\ee
The fields $V_{\alpha}$ and $V_{Q-\alpha}$ have the same conformal dimensions and represent
the same primary field, i.e. they are proportional to each other, and it follows from (\ref{twopf}) and  (\ref{twopfq})
that 
\be
V_{\alpha}=S(\alpha)V_{Q-\alpha}
\ee
\item
The three-point functions ${\cal C}(\alpha_1,\alpha_2,\alpha_3)$ for the values of $\alpha_i$ satisfying the relation
\be\label{conserva}
\alpha_1+\alpha_2+\alpha_3=Q-nb\, ,
\ee
are given by the Coulomb gas or screening integrals computed in \cite{Dotsenko:1984ad}
\be\label{df}
I_n(\alpha_1,\alpha_2,\alpha_3)=\left(b^4\gamma(b^2)\pi\mu\right)^n
{\prod_{j=1}^n\gamma(-jb^2)\over \prod_{k=0}^{n-1}[\gamma(2\alpha_1 b+kb^2)\gamma(2\alpha_2 b+kb^2)
\gamma(2\alpha_3 b+kb^2)]}\, ,
\ee
\end{enumerate}
where $\gamma(x)={\Gamma(x)\over \Gamma(1-x)}$.

The structure constants derived as the Coulomb gas integrals are denoted by ${\cal C}$ to distinguish from their values derived from the 
DOZZ formula.

The structure constant are related to the three-point functions by the relation:
\be\label{struc}
C^{\alpha_3}_{\alpha_1,\alpha_2}=C(\alpha_1,\alpha_2,Q-\alpha_3)\, .
\ee
Thus we derive:
\be\label{ccul1}
{\cal C}_{-b/2,\alpha}^{\alpha-b/2}=1\, ,
\ee
and
\be
{\cal C}_{-b/2,\alpha}^{\alpha+b/2}={\pi\mu b^4\gamma(b^2)\over \gamma(2\alpha b)\gamma(b^2-2\alpha b+2)}\, .
\ee
Now one can obtain the two-point function $S(\alpha)$  by the following trick \cite{Fateev:2000ik}.
Consider the auxiliary three-point function
\be\label{auxf}
\langle V_{\alpha}(x_1)V_{\alpha+b/2}(x_2)V_{-b/2}(z)\rangle\, .
\ee
Using the OPE
\be
V_{-b/2}V_{\alpha}={\cal C}_{-b/2,\alpha}^{\alpha-b/2}\left[V_{\alpha-b/2}\right]+{\cal C}_{-b/2,\alpha}^{\alpha+b/2}\left[V_{\alpha+b/2}\right]\, ,
\ee
one receives that in the limit $z\rightarrow x_1$ the three-point function (\ref{auxf}) 
takes the form:
 \be\label{stg1}
{\cal C}_{-b/2,\alpha}^{\alpha+b/2}S(\alpha+b/2)\, ,
\ee
 whereas in the limit $z\rightarrow x_2$, it is
\be\label{stg2}
{\cal C}_{-b/2,\alpha}^{\alpha-b/2}S(\alpha)\, .
\ee
Equating (\ref{stg1}) and (\ref{stg2}) we get that the two-point function $S(\alpha)$
satisfies the condition:
\be\label{reefcond}
{S(\alpha)\over S(\alpha+b/2)}={\cal C}_{-b/2,\alpha}^{\alpha+b/2}\, .
\ee
Solving (\ref{reefcond}) one derives:
\be\label{reflal}
S(\alpha)={\left(\pi\mu\gamma(b^2)\right)^{b^{-1}(Q-2\alpha)}\over b^2}{\Gamma(1-b(Q-2\alpha))\Gamma(-b^{-1}(Q-2\alpha))
\over \Gamma(b(Q-2\alpha))\Gamma(1+b^{-1}(Q-2\alpha))}\, .
\ee

We have all the necessary ingredients to compute classifying algebra: two-point function $S(\alpha)$
and vacuum component of the matrix of the modular transformation. Before to continue let us
recall that both of them can be conveniently written using ZZ function \cite{Zamolodchikov:2001ah}:
\be\label{zzfu}
W(\alpha)=-{2^{3/4}e^{3i\pi/2}(\pi\mu\gamma(b^2))^{-{(Q-2\alpha)\over 2b}}\pi(Q-2\alpha)\over
\Gamma(1-b(Q-2\alpha))\Gamma(1-b^{-1}(Q-2\alpha))}\, .
\ee
It can be easily shown that
\be\label{salpha}
{W(Q-\alpha)\over W(\alpha)}=S(\alpha)\, ,
\ee
and
\be\label{wsal}
W(Q-\alpha)W(\alpha)=S_{0\alpha}\, .
\ee
Recalling (\ref{fas}), $F_{\alpha}$ takes the form:
\be\label{falpha}
F_{\alpha}={S_{00}\over W(Q-\alpha)W(\alpha)}\, .
\ee
Combining (\ref{salpha}) and (\ref{falpha}) we obtain coefficients $\xi_{\alpha}$ for
the Liouville field theory:

\be\label{xil}
\xi^L_{\alpha}=\sqrt{S(\alpha)F(\alpha)}={\sqrt{S_{00}}\over W(\alpha)}\, .
\ee
Eq. (\ref{cpij}) implies:
\be\label{cfal}
C_{\alpha_1,\alpha_2}^{\alpha_3}F_{\alpha_3,0}\left[\begin{array}{cc}
\alpha_1&\alpha_1\\
\alpha_2 &\alpha_2 \end{array}\right]=W(0){W(\alpha_3)\over W(\alpha_1)W(\alpha_2)}\, .
\ee
As we explained in section 2, in formulae (\ref{xil}) and (\ref{cfal}) could appear a sign factor. But below we check that  it is absent here.

Let us compare  (\ref{cfal}) with the calculations in literature.
First of all recall  the calculations in \cite{Fateev:2000ik} for one of the momenta taking the degenerate value $\alpha_1=-{b\over 2}$.
The fusing matrix  can be computed using that conformal blocks with the degenerate primary $-{b\over 2}$ satisfy
the second order differential equation, which can be solved by the hypergeometric functions.
The fusion matrix is given by the transformation properties of the hypergeometric functions.
The fusion matrix   derived in this way we denote by  $F^{*}$ to distinguish from the values of the fusion matrix derived from the Ponsot-Teschner formula.
The corresponding values of $F^{*}$ are  \cite{Fateev:2000ik,Teschner:1995yf}:
\be\label{fatmat1}
F^{*}_{\alpha-b/2,0}\left[\begin{array}{cc}
-b/2&-b/2\\
\alpha &\alpha \end{array}\right]={\Gamma(2\alpha b-b^2)\Gamma(-1-2b^2)\over 
\Gamma(2\alpha b-2b^2-1)\Gamma(-b^2)}\, ,
\ee

\be\label{fatmat2}
F^{*}_{\alpha+b/2,0}\left[\begin{array}{cc}
-b/2&-b/2\\
\alpha &\alpha \end{array}\right]={\Gamma(2+b^2-2\alpha b)\Gamma(-1-2b^2)\over 
\Gamma(1-2\alpha b)\Gamma(-b^2)}\, .
\ee

Using ZZ function $W(\alpha)$ (\ref{zzfu})  one can compactly rewrite  (\ref{fatmat1}), (\ref{fatmat2}) as:

\be\label{fatmat1r}
F^{*}_{\alpha-b/2,0}\left[\begin{array}{cc}
-b/2&-b/2\\
\alpha &\alpha \end{array}\right]={W(0)\over W(-{b\over 2})}{W(\alpha- b/2)\over W(\alpha)}\, ,
\ee

\be\label{fatmat2r}
F^{*}_{\alpha+b/2,0}\left[\begin{array}{cc}
-b/2&-b/2\\
\alpha &\alpha \end{array}\right]={W(0)\over W(-{b\over 2})}{W(Q-\alpha- b/2)\over W(Q-\alpha)}\, .
\ee

Combining  (\ref{ccul1}), (\ref{reefcond}),  (\ref{salpha}), (\ref{fatmat1r}), (\ref{fatmat2r})   we obtain

\be\label{fzzz1}
{\cal C}_{-b/2,\alpha}^{\alpha-b/2}F^{*}_{\alpha-b/2,0}\left[\begin{array}{cc}
-b/2&-b/2\\
\alpha &\alpha \end{array}\right]={W(0)\over W(-{b\over 2})}{W(\alpha- b/2)\over W(\alpha)}\, ,
\ee

\be\label{fzzz2}
{\cal C}_{-b/2,\alpha}^{\alpha+b/2}F^{*}_{\alpha+b/2,0}\left[\begin{array}{cc}
-b/2&-b/2\\
\alpha &\alpha \end{array}\right]={W(0)\over W(-{b\over 2})}{W(\alpha+b/2)\over W(\alpha)}\, ,
\ee

in agreement with (\ref{cfal}).

Next we compute the left hand side of (\ref{cfal}) using DOZZ formula for structure constants \cite{Dorn:1994xn,Zamolodchikov:1995aa} and the explicit expression for the fusing matrix found in \cite{Ponsot:1999uf}.
It is instructive at the beginning to repeat the steps leading from (\ref{asseq})  to (\ref{cpij})
for the Liouville theory using the DOZZ formula.
Using the relation between three-point functions and OPE structure constant (\ref{struc})
the associativity condition of the OPE in the Liouville field theory  takes the form:

\bea\label{assope}
&&C(\alpha_4,\alpha_3,\alpha_s)C(Q-\alpha_s,\alpha_2,\alpha_1)F_{\alpha_s,\alpha_t}\left[\begin{array}{cc}
\alpha_3&\alpha_2\\
\alpha_4 &\alpha_1 \end{array}\right]=\\ \nonumber
&&=C(\alpha_4,\alpha_t,\alpha_1)C(Q-\alpha_t,\alpha_3,\alpha_2)F_{\alpha_t,\alpha_s}\left[\begin{array}{cc}
\alpha_1&\alpha_2\\
\alpha_4 &\alpha_3 \end{array}\right]\, .
\eea
Consider the limit $\alpha_t\rightarrow 0$ in (\ref{assope}).

From the DOZZ formula:
\bea\label{dozz}
&&C(\alpha_1,\alpha_2,\alpha_3)=\lambda^{(Q-\sum_{i=1}^3\alpha_i)/b}\times\\ \nonumber
&&{\Upsilon_b(b)\Upsilon_b(2\alpha_1)\Upsilon_b(2\alpha_2)\Upsilon_b(2\alpha_3)\over
\Upsilon_b(\alpha_1+\alpha_2+\alpha_3-Q)\Upsilon_b(\alpha_1+\alpha_2-\alpha_3)
\Upsilon_b(\alpha_2+\alpha_3-\alpha_1)\Upsilon_b(\alpha_3+\alpha_1-\alpha_2)}\, ,
\eea
where \be
\lambda=\pi\mu\gamma(b^2)b^{2-2b^2}
\ee

one can obtain  \cite{Teschner:2001rv}
\be\label{climit}
C(\alpha_2,\epsilon,\alpha_1)\simeq {2\epsilon S(\alpha_1)\over (\alpha_2-\alpha_1+\epsilon)
(\alpha_1-\alpha_2+\epsilon)}+
{2\epsilon \over (Q-\alpha_2+\alpha_1+\epsilon)
(\alpha_1+\alpha_2-Q+\epsilon)}\, .
\ee
The functions $\Upsilon_b(\alpha)$ and their properties leading to (\ref{climit}) are described in appendix B.

Using the reflection property
\be\label{reflector}
C(\alpha_3,\alpha_2,\alpha_1)=S(\alpha_3)C(Q-\alpha_3,\alpha_2,\alpha_1)\, ,
\ee
one receives in this limit, setting also $\alpha_1=\alpha_4$,  $\alpha_2=\alpha_3$
\be\label{csf}
C^2(\alpha_2,\alpha_1,\alpha_s)={4S(\alpha_1)S(\alpha_2)S(\alpha_s)\over S(0)}
{F_{0,\alpha_s}\left[\begin{array}{cc}
\alpha_1&\alpha_2\\
\alpha_1 &\alpha_2 \end{array}\right]\over
 {\rm lim}_{\epsilon\rightarrow 0}\epsilon^2F_{\alpha_s,\epsilon}\left[\begin{array}{cc}
\alpha_2&\alpha_2\\
\alpha_1 &\alpha_1 \end{array}\right]}\, .
\ee
It was shown in \cite{Teschner:2008qh} that the limit
\be\label{predel}
F''_{\alpha,0}\left[\begin{array}{cc}
\alpha_3&\alpha_2\\
\alpha_4 &\alpha_1 \end{array}\right]\equiv {\rm lim}_{\beta\rightarrow 0}\beta^2
F_{\alpha,\beta}\left[\begin{array}{cc}
\alpha_3&\alpha_2\\
\alpha_4 &\alpha_1 \end{array}\right]
\ee
exists
and satisfies the equation:
\be\label{limf}
F''_{\alpha,0}\left[\begin{array}{cc}
\alpha_2&\alpha_2\\
\alpha_1 &\alpha_1 \end{array}\right]F_{0,\alpha}\left[\begin{array}{cc}
\alpha_2&\alpha_1\\
\alpha_2 &\alpha_1 \end{array}\right]
={F_{\alpha_2}F_{\alpha_1}\over F_{\alpha}}\, .
\ee

Putting (\ref{limf}) in (\ref{csf}) one finally gets:
\be\label{cal12}
C(\alpha_1,\alpha_2,\alpha_s)F''_{\alpha_s,0}\left[\begin{array}{cc}
\alpha_1&\alpha_1\\
\alpha_2 &\alpha_2 \end{array}\right]=2W(0){W(Q-\alpha_s)\over W(\alpha_1)W(\alpha_2)}\, .
\ee
and
\be\label{cal1234}
C(\alpha_1,\alpha_2,\alpha_s)=2W(Q-\alpha_1){W(Q-\alpha_2)\over W(Q)W(\alpha_s)}F_{0,\alpha_s}\left[\begin{array}{cc}
\alpha_1&\alpha_2\\
\alpha_1 &\alpha_2 \end{array}\right]\, .
\ee
Here a sign factor could appear, but below we show that actually  (\ref{cal12}) and (\ref{cal1234})  hold without it.
Recalling the relation (\ref{struc}) and (\ref{reflector}) we obtain (\ref{cfal}).
The emergence of the factor 2 will be explained below.
This derivation also explains that the double pole in the fusing matrix $F_{\alpha,0}\left[\begin{array}{cc}
\alpha_1&\alpha_1\\
\alpha_2 &\alpha_2 \end{array}\right]$ is related to the simple pole
in the DOZZ formula.

One can  compute  the limit (\ref{predel})  also directly\footnote{See for similar calculations also
 \cite{Petkova:2009pe}.}.
Recall that the boundary three-point function is given by \cite{Ponsot:2001ng}
\be
C^{\sigma_3\sigma_2\sigma_1}_{Q-\beta_3\beta_2\beta_1}=
C^{\sigma_3\sigma_2\sigma_1}_{\beta_3|\beta_2\beta_1}={g^{\sigma_3\sigma_1}_{\beta_3}
\over g^{\sigma_3\sigma_2}_{\beta_2}g^{\sigma_2\sigma_1}_{\beta_1}}
F_{\sigma_2\beta_3}\left[\begin{array}{cc}
\beta_2&\beta_1\\
\sigma_3 &\sigma_1 \end{array}\right]\, ,
\ee
where
\be\label{gsigbet}
g^{\sigma_3\sigma_1}_{\beta}=
\lambda^{\beta/2b}{\Gamma_b(Q)\Gamma_b(Q-2\beta)\Gamma_b(2\sigma_1)\Gamma_b(2Q-2\sigma_3)\over
\Gamma_b(2Q-\beta-\sigma_1-\sigma_3)\Gamma_b(\sigma_1+\sigma_3-\beta)
\Gamma_b(Q-\beta+\sigma_1-\sigma_3)\Gamma_b(Q-\beta+\sigma_3-\sigma_1)}\, .
\ee
The function $\Gamma_b(x)$ is described in appendix B.

Therefore the fusing matrix can be expressed as
\be
F_{\sigma_2\beta_3}\left[\begin{array}{cc}
\beta_2&\beta_1\\
\sigma_3 &\sigma_1 \end{array}\right]={g^{\sigma_3\sigma_2}_{\beta_2}g^{\sigma_2\sigma_1}_{\beta_1}
\over g^{\sigma_3\sigma_1}_{\beta_3}}C^{\sigma_3\sigma_2\sigma_1}_{Q-\beta_3\beta_2\beta_1}\, .
\ee

On the other side $C^{\sigma_3\sigma_2\sigma_1}_{Q-\beta_3\beta_2\beta_1}$ has a pole
with residue 1 if $\beta_1+\beta_2-\beta_3=0$.
Therefore using the invariance of the fusing matrix w.r.t.  to the inversions $\alpha_i\rightarrow Q-\alpha_i$
one can write for the corresponding residue of the fusion matrix 
\be
F'_{\sigma_2,0}\left[\begin{array}{cc}
\beta_1&\beta_1\\
\sigma_1 &\sigma_1 \end{array}\right]=F'_{\sigma_2,Q}\left[\begin{array}{cc}
Q-\beta_1&\beta_1\\
\sigma_1 &\sigma_1 \end{array}\right]={g^{\sigma_1\sigma_2}_{Q-\beta_1}g^{\sigma_2\sigma_1}_{\beta_1}
\over g^{\sigma_1\sigma_1}_{Q}}\, .
\ee

Using the explicit  expressions (\ref{gsigbet})  for $g^{\sigma_2\sigma_1}_{\beta_1}$, the DOZZ
formula (\ref{dozz}) for structure constants and the properties of the functions $ \Gamma_b(x)$, $\Upsilon_b(x)$ reviewed in appendix B, 
 it is easy to compute that

\be\label{gsig}
g^{\sigma_1\sigma_2}_{Q-\beta_1}g^{\sigma_2\sigma_1}_{\beta_1}
=2^{1/4}e^{-3i\pi/2}{2\pi W(Q-\sigma_1)W(Q-\sigma_2)\over
W(\beta_1)}{1\over C(\sigma_1,\sigma_2,\beta_1)}\, .
\ee
Using the properties of the functions $ \Gamma_b(x)$,  reviewed in appendix B,
one can compute the limit
\be
{\rm lim}_{\beta_3\rightarrow Q}{1\over g^{\sigma_1\sigma_1}_{\beta_3}}\, ,
\ee
and obtain that it has simple pole with the residue
\be\label{resid}
{2^{-1/4}e^{3i\pi/2}W(0)\over \pi W(\sigma_1)W(Q-\sigma_1)}\, .
\ee

Combining  (\ref{gsig}) and (\ref{resid}) we again derive (\ref{cal12}).

Some comments are in order at this point:
\begin{enumerate}
\item
This derivation shows that the fusing matrix element $F_{\sigma_2,0}\left[\begin{array}{cc}
\beta_1&\beta_1\\
\sigma_1 &\sigma_1 \end{array}\right]$
indeed has double pole: one degree comes from the pole of the three-point function
$C^{\sigma_3\sigma_2\sigma_1}_{0,Q-\beta_1,\beta_1}$  and the second
from the pole of the ${1\over g^{\sigma_1\sigma_1}_Q}$.
\item
We have shown that (\ref{cfal}) or (\ref{cal12}) indeed always holds with the understanding that
in the case of the singular behavior one should take the coefficients of the leading singularities.
\item
Note that (\ref{cal12}) evidently satisfies the reflection property (\ref{reflector})
since the fusing matrix is invariant under the inversions $\alpha\rightarrow Q-\alpha$.
\item
Let us explain the emergence of the factor 2 in (\ref{cal12}).
We have seen that the formula (\ref{cal12}), derived by using the DOZZ formula for structure constant and Ponsot-Teschner  (PT)
formula for the fusing matrix has additional factor $2$ compared to formulas (\ref{cfal}), (\ref{fzzz1}), (\ref{fzzz2})
using the values of the structure constant derived as the Coulomb gas integrals and fusing matrix computed via the differential equations
for the conformal blocks. The derivation of the formula (\ref{cal12}) via the limiting procedure (\ref{csf})-(\ref{cal12})
indicates that the factor $2$ originates from the coefficient $2$ in formula (\ref{climit}). Point is that as the formula (\ref{climit}) shows, the  two-point functions,
derived from the DOZZ formula as residue of the pole in the limit $\alpha_3\rightarrow 0$,  are twice  the two-point functions (\ref{twopf}) and (\ref{twopfq}),
derived in the Coulomb gas approach. Thus the states in the theory reconstructed  from the DOZZ formula have twice the normalization
of the fields used in the calculations leading to  (\ref{cfal}), (\ref{fzzz1}), (\ref{fzzz2}). This is the reason for emergence of the factor $2$ in (\ref{cal12}).
\item
One can ask, what happens if one tries to compute the left hand side of  formulae (\ref{fzzz1}), (\ref{fzzz2}) from the DOZZ and PT formulae.
First of all let us recall that, as noted in \cite{Zamolodchikov:1995aa}, when the momenta $\alpha_i$  satisfy the relation (\ref{conserva}),
the DOZZ formula has a pole with the residue equals to the Coulomb gas integrals (\ref{df}):
\be
{\rm res}_{\alpha_1+\alpha_2+\alpha_3=Q}C(\alpha_1,\alpha_2.\alpha_3)=1\, ,
\ee
and
\be
{\rm res}_{\alpha_1+\alpha_2+\alpha_3=Q-nb}C(\alpha_1,\alpha_2.\alpha_3)=I_n(\alpha_1,\alpha_2,\alpha_3)\, .
\ee

But strictly speaking this is true only for the non-degenerate values of the momenta.
For the degenerate values (\ref{deger}), as we see from the DOZZ formula, it may happen that additionally to  the first vanishing term in the denominator, we have 
two more vanishing terms, one  in the denominator and another one   in the numerator. This makes the limiting procedure ambiguous and  can bring to 
the  values of the residue twice as the Coulomb gas results. 

Consider the values of the momenta appearing  in formulae (\ref{fzzz1}), (\ref{fzzz2}): $\alpha_1=\alpha$, $\alpha_2=-{b\over 2}$, 
$\alpha_3=Q-\alpha\pm {b\over 2}$. For these $\alpha_i$  the  DOZZ  formula develops pole, and  the matrix
$F''$ defined in (\ref{predel}), vanishes.
If now we set $\alpha_2=-{b\over 2}+\delta$ and consider the limit $\delta\rightarrow 0$,  we obtain 
\be
{\rm lim}_{\delta\rightarrow 0}\;\delta \; C\left(\alpha,-{b\over 2}+\delta,Q-\alpha\mp {b\over 2}\right)=2{\cal C}_{-b/2,\alpha}^{\alpha\pm b/2}\, ,
\ee

\be
{\rm lim}_{\delta\rightarrow 0}{1\over \delta} F''_{\alpha\pm b/2,0}\left[\begin{array}{cc}
-b/2+\delta&-b/2+\delta\\
\alpha &\alpha \end{array}\right]=F^{*}_{\alpha\pm b/2,0}\left[\begin{array}{cc}
-b/2&-b/2\\
\alpha &\alpha \end{array}\right]\, .
\ee
On the other hand it was suggested in \cite{Harlow:2011ny} a limiting procedure reproducing the Coulomb gas values:
\be
{\rm lim}_{\delta\rightarrow 0}\left[{\rm lim}_{\epsilon\rightarrow 0}\;\epsilon\; C\left(\alpha,-{b\over 2}+\delta,Q-\alpha\mp {b\over 2}-\delta+\epsilon\right)\right]=
{\cal C}_{-b/2,\alpha}^{\alpha\pm b/2}\, .
\ee

But this procedure brings to the factor $2$ in the fusion matrix:

\be
{\rm lim}_{\delta\rightarrow 0}\left[{\rm lim}_{\epsilon\rightarrow 0}{1\over \epsilon} F''_{\alpha\pm b/2+\delta-\epsilon,0}\left[\begin{array}{cc}
-b/2+\delta&-b/2+\delta\\
\alpha &\alpha \end{array}\right]\right]=2F^{*}_{\alpha\pm b/2,0}\left[\begin{array}{cc}
-b/2&-b/2\\
\alpha &\alpha \end{array}\right]\, .
\ee
In any case we are in agreement (\ref{cal12}).
\end{enumerate}

Having demonstrated that (\ref{cfal}) holds in the Liouville field theory we can use the formulae of the section 2
for the defects and boundaries. This will enable us to rederive and write down the formulas for
D-branes in the Liouville field theory derived in \cite{Fateev:2000ik} and \cite{Zamolodchikov:2001ah},
and for defects and permutation branes derived in \cite{Sarkissian:2009aa} in the simple and elegant way.

With $j=-{b\over 2}$,  $i=\alpha$, and $k=\alpha \pm b/2$, the equations (\ref{nbk}) and (\ref{nbkc}) take the forms:
\be\label{psik}
\Psi(\alpha)\Psi(-b/2)=\Psi(\alpha-b/2)+\Psi(\alpha+b/2)\, ,
\ee
and

\be\label{tsik}
{W(-b/2)\over W(0)}A\Lambda(\alpha)=\Lambda(\alpha-b/2)+\Lambda(\alpha+b/2)
\ee
correspondingly.

The solution of the equations (\ref{psik}) and (\ref{tsik}) are
\be
\Psi_{m,n}(\alpha)={\sin(\pi m b^{-1}(2\alpha- Q))\sin(\pi nb (2\alpha-Q))\over \sin(\pi mb^{-1} Q)\sin(\pi nb Q)}={S_{m,n\,\alpha}\over S_{m,n\,0}}\, ,
\ee

and
\be\label{funcla}
\Lambda_s(\alpha)=2^{1/2}\cosh(2\pi s(2\alpha-Q))\, ,
\ee
with
\be\label{rela}
2\cosh 2\pi bs=A{W(-b/2)\over W(0)}
\ee
respectively.

Using equations of  section 2  we obtain one-point functions for ordinary branes
\be
U_{m,n}(\alpha)=\Psi_{m,n}(\alpha){W(0)\over W(\alpha)}\, ,
\ee
 permutation branes on $N$-fold product
\be
U_{{\cal P}m,n}^N(\alpha)=\Psi_{m,n}(\alpha)\left({W(0)\over W(\alpha)}\right)^N\, ,
\ee
and defects
\be
D_{m,n}(\alpha)=\Psi_{m,n}(\alpha){S_{00}\over S_{0\alpha}}\, .
\ee

Using (\ref{ub}) one derives boundary state coefficients for ordinary branes:
\be
B_{m,n}(\alpha)={S_{m,n\,\alpha}\over W(\alpha)}\, ,
\ee
permutation branes on $N$-fold product
\be
B_{{\cal P}m,n}^N(\alpha)={S_{m,n\,\alpha}\over W^N(\alpha)}\, ,
\ee
and defects
\be
{\cal D}_{m,n}(\alpha)={S_{m,n\,\alpha}\over S_{0\alpha}}\, .
\ee

For the continuous family  one gets similarly,
using (\ref{nbkc}),  (\ref{nbkc2}), and  (\ref{nbkc3}),  boundary state coefficients for ordinary branes
\be
B_{s}(\alpha)={\Lambda_s(\alpha)\over W(\alpha)}\, ,
\ee
 permutation branes on $N$-fold product
\be
B_{{\cal P}s}^N(\alpha)={\Lambda^{(N)}_{s{\cal P}}(\alpha)\over W^N(\alpha)}\, ,
\ee
and defects
\be
{\cal D}_{s}(\alpha)={\Lambda^{(2)}_{s{\cal P}}(\alpha)\over S_{0\alpha}}\, .
\ee
$\Lambda^{(N)}_{s{\cal P}}(\alpha)$ is again given by the function (\ref{funcla}), but the relation
(\ref{rela}) now takes the form
\be\label{relaa}
2\cosh 2\pi bs=A\left({W(-b/2)\over W(0)}\right)^N\, .
\ee

\section{Toda field theory}
Recall  some facts on Toda field theory  \cite{Fateev:2007ab}. The action of the $sl(n)$
conformal Toda field theory on a two-dimensional surface with metric $g_{ab}$ and associated to it scalar curvature
$R$ has the form
\be
{\cal A}=\int\left({1\over 8\pi}g_{ab}(\partial_a\varphi\partial_b \varphi)+\mu \sum_{k=1}^{n-1}e^{b(e_k\varphi)}+
{(Q,\varphi)\over 4\pi}R\right)\sqrt{g}d^2x\, .
\ee
Here $\varphi$ is the two-dimensional $(n-1)$ component scalar field $\varphi=(\varphi_1\ldots \varphi_{n-1})$:
\be
\varphi=\sum_i^{n-1}\varphi_ie_i\, ,
\ee
where vectors $e_k$ are the simple roots of the Lie algebra $sl(n)$,
$b$ is the dimensionless coupling constant, $\mu$ is the scale parameter called the cosmological constant and $(e_k,\varphi)$
denotes the scalar product.

If the background charge $Q$ is related with the parameter $b$ as
\be
Q=\left(b+{1\over b}\right)\rho\, ,
\ee
where $\rho$ is the Weyl vector , then the theory
is conformally invariant.
The Weyl vector is
\be
\rho={1\over 2}\sum_{e>0}e=\sum_i^{n-1}\omega_i\, ,
\ee
where $\omega_i$ are fundamental weights, such that $(\omega_i, e_j)=\delta_{ij}$.

Conformal Toda field theory possesses  higher-spin symmetry: there are $n-1$ holomorphic currents $W^k(z)$ with the
spins $k=2,3,\ldots n$. The currents $W^k(z)$ form closed $W_n$ algebra, which contains as subalgebra
the Virasoro algebra with the central charge
\be
c=n-1+12Q^2=(n-1)(1+n(n-1)(b+b^{-1}))\, .
\ee

Primary fields of conformal Toda field theory are the exponential field parameterized by a $(n-1)$
component vector parameter $\alpha$, $\alpha=\sum_i^{n-1}\alpha_i\omega_i$,
\be
V_{\alpha}=e^{(\alpha,\varphi)}\, .
\ee

They have the simple OPE with the currents $W^k(z)$.  Namely,
\be
W^k(\xi)V_{\alpha}(z,\bar{z})={w^{(k)}(\alpha)V_{\alpha}(z,\bar{z})\over
(\xi-z)^k}\, .
\ee
The quantum numbers $w^{(k)}(\alpha)$ possess the symmetry under the action of the Weyl
group ${\cal W}$ of the algebra $sl(n)$:
\be\label{wsa}
w^{(k)}(\alpha)=w^{(k)}(Q+\hat{s}(\alpha-Q)),\,\,\, \hat{s}\in  {\cal W}\, .
\ee
In particular
\be
w^{(2)}(\alpha)=\Delta(\alpha)={(\alpha, 2Q-\alpha)\over 2}
\ee
is the conformal dimension of the field $V_{\alpha}$.
Eq. (\ref{wsa})  implies that the fields related via the action of the Weyl group should coincide up to a multiplicative factor. Indeed we have \cite{Fateev:2001mj}
\be
R_{ \hat{s}}(\alpha)V_{Q+\hat{s}(\alpha-Q)}=V_{\alpha}\, ,
\ee
where $R_{ \hat{s}}(\alpha)$ is the reflection amplitude
\be
R_{ \hat{s}}(\alpha)={A(Q+\hat{s}(\alpha-Q))\over A(\alpha)}\, ,
\ee
\be
A(\alpha)=(\pi\mu\gamma(b^2))^{{(\alpha-Q,\rho)\over b}}{2\pi b \sqrt{\Xi}\over \prod_{e>0}\Gamma(1-b(\alpha-Q,e))
\Gamma(-b^{-1}(\alpha-Q,e))}\, ,
\ee
where
\be\label{ksi}
\Xi=i^{n-1}\sqrt{{\rm det}C}{1\over |{\cal W}|}\, ,
\ee
and  $C$ is the Cartan matrix.
Two-point functions in Toda field theory are
\be\label{twopft}
\langle V_{\alpha}(z_1,\bar{z}_1)V_{\alpha^*}(z_2,\bar{z}_2)\rangle={R(\alpha)\over (z_1-z_2)^{4\Delta_{\alpha}}
(\bar{z}_1-\bar{z}_2)^{4\Delta_{\alpha}}}\, ,
\ee
where $R(\alpha)$ is the maximal reflection amplitude defined as
\be
R(\alpha)={A(2Q-\alpha)\over A(\alpha)}\, ,
\ee
and $\alpha^*$ is defined by
\be\label{conj}
(\alpha,e_k)=(\alpha^*,e_{n-k})\, .
 \ee

The representations which appear  in the spectrum of $sl(n)$ Toda field theory have momenta
\be
\alpha\in Q+i\sum_i^{n-1} p_i\omega_i\, ,
\ee
where $p_i$ are real.

To describe degenerate representations it is useful to write $\alpha$ as
\be
\alpha=Q+\nu\, .
\ee
Degenerate  representations appear at  the momentum $\nu$ satisfying the condition
\be\label{mbrs}
-(\nu,e)=rb+{s\over b}\, ,
\ee
where $e$ is a root and $r,s\in \mathbb{Z}_{+}$ .
Without loss of generality we can classify semi-degenerate representations by a collection of simple roots
${\cal I}$ for which the equation is satisfied:
\be\label{mbrsi}
-(\nu,e_i)=rb+{s\over b}\hspace{1cm} i\in {\cal I}\, .
\ee

Fully degenerate  representations appear when ${\cal I}$  consists of all the simple roots.
It is easy to show that for fully degenerate representations $\alpha$ takes the form:
\be
\alpha_{R_1|R_2}=-b\lambda_1-{1\over b}\lambda_2\, ,
\ee
where $\lambda_1$ and $\lambda_2$ are the highest weights correspondning to   irreducible
representations $R_1$  and $R_2$ of $sl(n)$.

The identity representation, as in the Liouville case before, belongs to the set of the
 fully degenerate representations.

To characterize generic semi-degenerate representations we need more notations.
Denote by $\Delta_{\cal I}$ subsystem of roots which are linear combinations of the simple roots in ${\cal I}$,
and by $\rho_{\cal I}$ restricted Weyl vector as half sum of the positive roots in  $\Delta_{\cal I}$.
For semi-degenerate representations $\nu$ takes the form
\be
\nu_{\tilde{\nu},R_1,R_2}=\tilde{\nu}-(\rho_{\cal I}+\lambda_1)b-(\rho_{\cal I}+\lambda_2)/b\, ,
\ee
where $\tilde{\nu}$ is continuous component of the $\nu$ in the direction orthogonal to simple roots
in ${\cal I}$, and $\lambda_1$ and $\lambda_2$ are the highest weights correspomding to irreducible
representations $R_1$  and $R_2$ of the Lie algebra built from  $\Delta_{\cal I}$.
The elements of the matrix of the modular transformation have been computed in  \cite{Drukker:2010jp}
and given by the following expressions:
\be
S_{\beta\alpha}=\Xi\sum_{\omega\in W}\epsilon(\omega)e^{2\pi i(\omega(\beta-Q),\alpha-Q)}\, ,
\ee

\be
S_{R_1|R_2,\alpha}=\chi_{R_1}(e^{2\pi i b (Q-\alpha)})\chi_{R_2}(e^{2\pi i b^{-1} (Q-\alpha)})S_{0\alpha}\, ,
\ee

\be
S_{0\alpha}=\Xi\prod_{e>0}4\sin(\pi b (\alpha-Q,e))\sin(-{\pi\over b} (\alpha-Q,e))\, ,
\ee

\bea
&&S_{\tilde{\nu}R_1|R_2,\alpha}=\Xi\sum_{\tilde{\omega}\in W/W_{\cal I}}\epsilon(\omega)
e^{2\pi i(\tilde{\omega}(\tilde{\mu}),\alpha-Q)}
\chi_{R_1}(e^{2\pi ib\tilde{\omega}^{-1} (Q-\alpha)})\times \\ \nonumber
&&\chi_{R_2}(e^{2\pi ib^{-1}\tilde{\omega}^{-1} (Q-\alpha)})
\prod_{e\in \Delta^+_{\cal I}}4\sin(\pi b (\alpha-Q,\tilde{\omega}(e)))\sin(-{\pi\over b} (\alpha-Q,\tilde{\omega}(e)))\, .
\eea

$\chi_R(e^x)$ are the Weyl characters:
\be
\chi_R(e^x)={\sum_{\omega\in W}\epsilon(\omega)e^{(\omega(\rho+\lambda),x)}\over
\sum_{\omega\in W}\epsilon(\omega)e^{(\omega(\rho),x)}}
\ee
and $\Xi$ is defined by (\ref{ksi}).

Note that as in the Liouville field theory in the Toda field theory holds the relation as well
\be
A(\alpha)A(2Q-\alpha)=S_{0\alpha}\, .
\ee

Recalling (\ref{fas}),  we are ready to compute the coefficients $\xi_{\alpha}$ and $\eta_{\alpha}$ in the Toda field theory:
\be
\xi^T_{\alpha}=\epsilon_{\alpha}\sqrt{{A(2Q-\alpha)\over A(\alpha)}{S_{00}\over A(\alpha)A(2Q-\alpha)}}=\epsilon_{\alpha}{\sqrt{S_{00}}\over A(\alpha)}\, ,
\ee
\be
\eta^T_{\alpha}=\epsilon_{\alpha}\sqrt{{A(2Q-\alpha)\over A(\alpha)}{A(\alpha)A(2Q-\alpha)\over S_{00}}}=\epsilon_{\alpha}{A(2Q-\alpha)\over \sqrt{S_{00}}}\, .
\ee
Here $\epsilon_{\alpha}$ denotes a possible sign factor.

Therefore one has  in the Toda field theory
\be\label{cfat1}
C_{\alpha_1,\alpha_2,\mu\bar{\mu}}^{\alpha_3}={\epsilon_{\alpha_1}\epsilon_{\alpha_2}\over\epsilon_0\epsilon_{\alpha_3}} {A(2Q-\alpha_1)A(2Q-\alpha_2)\over A(2Q)A(2Q-\alpha_3)}F_{0,\alpha_3}\left[\begin{array}{cc}
\alpha_1&\alpha_2\\
\alpha_1 &\alpha_2^* \end{array}\right]_{00}^{\bar{\mu}\mu}\, .
\ee
Here $\mu$ and $\bar{\mu}$ label multiplicity of the representation $\alpha_3$ appearing in the fusion of  $\alpha_1$ and $\alpha_2$.
Eq. (\ref{cfat1}) implies:
\be\label{cfat}
\sum_{\mu\bar{\mu}}C_{\alpha_1,\alpha_2,\mu\bar{\mu}}^{\alpha_3}F_{\alpha_3,0}\left[\begin{array}{cc}
\alpha_1^*&\alpha_1\\
\alpha_2 &\alpha_2 \end{array}\right]^{00}_{\bar{\mu}\mu}={\epsilon_{\alpha_1}\epsilon_{\alpha_2}\over\epsilon_0\epsilon_{\alpha_3}}{A(0)A(\alpha_3)\over A(\alpha_1)A(\alpha_2)}N^{\alpha_3}_{\alpha_1\alpha_2}\, .
\ee
Some comments are in order at this point.
\begin{enumerate}
\item
Presently we have no closed expressions for fusing matrices and structure constants in the Toda field theory, and
cannot verify the expression (\ref{cfat1}) fully as we have done in the Liouville field theory. But in the absence of these expressions,
the formula  (\ref{cfat1}) can help to draw many conclusions  on different aspects of the Toda field theory.
\item
Actually we can use equation (\ref{cfat}) only for  $\alpha_1$, $\alpha_2$ and $\alpha_3$ possessing finite fusion multiplicity.
This is always true for important for us case of the degenerate representations.
\item
 In the Toda field theory one has also analogue of the relations (\ref{struc}) and (\ref{reflector})
in the Liouville field theory. In the  Toda field theory they read:
\be\label{struct}
C^{\alpha_3}_{\alpha_1,\alpha_2}=C(\alpha_1,\alpha_2,2Q-\alpha_3)
\ee
and
\be\label{reflectort}
C(\alpha_3^*,\alpha_2,\alpha_1)=R(\alpha_3)C(2Q-\alpha_3,\alpha_2,\alpha_1)\, .
\ee

It is easy to see that the relation (\ref{cfat1}) is in agreement  with (\ref{struct}) and (\ref{reflectort}),
observing that:

a) the fusing matrix is invariant under the Weyl reflections of the primaries, since they do not change the conformal dimensions,
and therefore it is invariant under the replacement $\alpha_i^*\rightarrow 2Q-\alpha_i$
of any of its variables,
and

b)  using the definition (\ref{conj}) one can prove that the function $A(\alpha)$   is the same for  $\alpha$ and
 $\alpha^*$
\be
A(\alpha)=A(\alpha^*)\, .
\ee
We assume that possible sign factors satisfy $\epsilon_{\alpha}=\epsilon_{\alpha^*}=\epsilon_{2Q-\alpha}$.

\item
It was computed  in \cite{Fateev:2010za}  that for $sl(3)$  Toda field theory

\be\label{ribfa}
C^{\alpha-bh}_{-b\omega_1,\alpha}F_{\alpha-bh,0}\left[\begin{array}{cc}
\alpha^* &\alpha \\
-b\omega_1 &-b\omega_1 \end{array}\right]=-{\Gamma(-2-3b^2)\over \Gamma(-b^2)}{\pi\mu\over \gamma(-b^2)}{A(\alpha-bh)\over A(\alpha)}\, ,
\ee

where $h\in H_{\omega_1}$  and  $H_{\omega_1}=\{\omega_1,\omega_2-\omega_1,-\omega_1\}$.

It is easy to show that for $sl(3)$  Toda field theory

\be
-{\Gamma(-2-3b^2)\over \Gamma(-b^2)}{\pi\mu\over \gamma(-b^2)}={A(0)\over A(-b\omega_1)}\, .
\ee

Recalling that for this case there are no multiplicities we have perfect  agreement with (\ref{cfat}). We also  see that for this case (\ref{cfat}) satisfied without any sign factor.
\item
All calculations leading  to (\ref{cfat1}), (\ref{cfat}) and (\ref{ribfa}) are performed in the Coulomb gas approach.
Calculations using exact expressions for the structure constants and fusing matrix, still unknown in the Toda field theory, may 
bring to the modifications similar to what we encountered in the Liouville field theory.
\end{enumerate}
The degenerate fields have in their OPE with general  primary $V_{\alpha}$ only finite number
of primaries $V_{\alpha'}$
\be
V_{-b\lambda_1-{1\over b}\lambda_2}V_{\alpha}=\sum_{s,p}C^{\alpha'_{sp}}_{-b\lambda_1-{1\over b}\lambda_2,\alpha}
V_{\alpha'_{sp}}\, ,
\ee
where $\alpha'_{sp}=\alpha-bh_s^{\lambda_1}-b^{-1}h_p^{\lambda_2}$.
$h_s^{\lambda_1}$ are weights of the representation $\lambda_1$:
\be
h_s^{\lambda_1}=\lambda_1-\sum_1^{n-1}s_ie_i\, ,
\ee
where $s_i$  are some non-negative integers.

Given the relation (\ref{cfat})
 we can write down the Cardy-Lewellen equations (\ref{clmul}) for Toda field theory
when  one of the primaries, say $j$, taken the degenerate one, using general
formalism developed in section 2.

Eq.  (\ref{nbk})  in Toda field theory takes the form:

\be\label{psiktsik}
\Psi(\alpha)\Psi(-b\omega_k)=\sum_s\Psi(\alpha-bh_s^{\omega_k})\, .
\ee

The solution of the equation  (\ref{psiktsik}) is given as in the rational conformal field theory
by the relation of elements of the matrix of the modular transformation:
\be
\Psi_{\lambda_1|\lambda_2}(\alpha)={S_{R_1|R_2,\alpha}\over S_{R_1|R_2,0}}\, .
\ee

Continuing  as in the previous  sections we obtain discrete family of the boundary state coefficients for ordinary branes, permutation
branes and defects:
\be
B_{R_1|R_2}(\alpha)={S_{R_1|R_2,\alpha}\over A(\alpha)}\epsilon_{\alpha}\, ,
\ee
\be
B^N_{{\cal P}\, R_1|R_2}(\alpha)={S_{R_1|R_2,\alpha}\over A^N(\alpha)}\epsilon_{\alpha}^N\, ,
\ee
\be
{\cal D}_{R_1|R_2}(\alpha)={S_{R_1|R_2,\alpha}\over S_{0\alpha}}\, .
\ee

The continuous family eq.  (\ref{nbkc}) takes the form:
\be\label{psiktsikc}
\Lambda(\alpha)A{A(-b\omega_k)\over A(0)}=\sum_s\Lambda(\alpha-bh_s^{\omega_k})\, .
\ee
The equation (\ref{psiktsikc})  as before can be solved by the elements of the matrix of modular transformation
corresponding to non-degenerate and semi-degenerate representations:
\be\label{lbet}
\Lambda_{\beta}(\alpha)=S_{\beta\alpha}\, ,
\ee
\be\label{lmu}
\Lambda_{\tilde{\mu}R_1|R_2}(\alpha)=S_{\tilde{\mu}R_1|R_2,\alpha}\, .
\ee
Dividing  (\ref{lbet}) and (\ref{lmu}) by $A(\alpha)/\epsilon_{\alpha}$, $A^N(\alpha)/\epsilon_{\alpha}^N$ and $S_{0\alpha}$,
we obtain ordinary branes, permutation branes and defects correspondingly.

\vspace{3cm}

\noindent {\bf Acknowledgments} \\[1pt]
This work was partially supported by the Research project 11-1c258 of
the State Committee of Science of Republic of Armenia and ANSEF hepth-2774 grant.

I also thank the High Energy Section of the Abdus Salam ICTP, Trieste,  where part of this work was done.

It is also pleasure to thank Valentina Petkova and Sylvain Ribault for useful comments.
\newpage
\appendix
\section{Properties of the fusing matrix}
Here we analyze various consequences of the pentagon equation \cite{Moore:1988qv,Moore:1989vd,Behrend:1999bn}:
\bea\label{pent1a}
\sum_{s,\beta_2, t_2,t_3}F_{p_2,s}\left[\begin{array}{cc}
j&k\\
p_1&b \end{array}\right]^{\beta_2t_3}_{\alpha_2\alpha_3}F_{p_1,l}\left[\begin{array}{cc}
i&s\\
a&b \end{array}\right]^{\gamma_1t_2}_{\alpha_1\beta_2}F_{s,r}\left[\begin{array}{cc}
i&j\\
l&k \end{array}\right]^{u_2u_3}_{t_2t_3}=\\ \nonumber
\sum_{\beta_1}
F_{p_1,r}\left[\begin{array}{cc}
i&j\\
a&p_2 \end{array}\right]^{\beta_1u_3}_{\alpha_1\alpha_2}F_{p_2,l}\left[\begin{array}{cc}
r&k\\
a&b \end{array}\right]^{\gamma_1u_2}_{\beta_1\alpha_3}\, .
\eea
First of all let us review some important  properties of the fusing matrix.

Fusing matrix  possesses the following symmetry properties \cite{Moore:1988qv}
\be\label{symmmul}
F_{p,q}\left[\begin{array}{cc}
k&j\\
i&l \end{array}\right]_{ab}^{cd}= F_{p^*,q}\left[\begin{array}{cc}
j&k\\
l^*&i^* \end{array}\right]_{ba}^{cd}= F_{p,q^*}\left[\begin{array}{cc}
i^*&l\\
k^*&j \end{array}\right]_{ab}^{dc}=F_{p^*,q^*}\left[\begin{array}{cc}
l&i^*\\
j^*&k \end{array}\right]_{ba}^{dc}\, .
\ee
Next we need to know behavior of the fusing matrix when one of the entries is the identity \cite{Behrend:1999bn,Fuchs:2002cm}:
\be\label{10}
F_{c,p}\left[\begin{array}{cc}
i&0\\
b&a \end{array}\right]^{\beta t}_{\alpha_1\alpha_2}=\delta_{pi}\delta_{ac}\delta_{\alpha_20}\delta_{t0}\delta_{\alpha_1\beta}\, ,
\ee
\be\label{20}
F_{c,p}\left[\begin{array}{cc}
0&j\\
b&a \end{array}\right]^{\beta t}_{\alpha_1\alpha_2}=\delta_{pj}\delta_{bc}\delta_{\alpha_10}\delta_{t0}\delta_{\alpha_2\beta}\, ,
\ee
\be\label{30}
F_{c,p}\left[\begin{array}{cc}
i&j\\
b&0 \end{array}\right]^{\beta t}_{\alpha_1\alpha_2}=\delta_{pb}\delta_{jc}\delta_{\alpha_1t}\delta_{\beta 0}\delta_{\alpha_20}\, ,
\ee
\be\label{40}
F_{c,p}\left[\begin{array}{cc}
i&j\\
0&a \end{array}\right]^{\beta t}_{\alpha_1\alpha_2}=\delta_{pa^*}\delta_{ci^*}\delta_{\alpha_2t}\delta_{\beta 0}\delta_{\alpha_10}\, .
\ee
The  equations (\ref{symmmul})  and  (\ref{40}) in some models hold only up to some  sign factors. 
Here for the sake of simplicity  we do not consider these factors, which after all do not change the main statements of the paper.

Now we are ready to derive the necessary relations.

Setting in (\ref{pent1a})  $p_1=0$, implying also $i=a$, $s=b^*$, $j^*=p_2$, $\beta_2=0$, $t_3=\alpha_3$,
$\alpha_1=0$, $\alpha_2=0$
one obtains:

\be\label{pent2a}
\sum_{t_2}F_{0,l}\left[\begin{array}{cc}
a&b^*\\
a&b \end{array}\right]^{\gamma_1t_2}_{00}F_{b^*,r}\left[\begin{array}{cc}
a&j\\
l&k \end{array}\right]^{u_2u_3}_{t_2\alpha_3}=\sum_{\beta_1}F_{0,r}\left[\begin{array}{cc}
a&j\\
a&j^* \end{array}\right]^{\beta_1u_3}_{00}
F_{j^*,l}\left[\begin{array}{cc}
r&k\\
a&b \end{array}\right]^{\gamma_1u_2}_{\beta_1\alpha_3}\, .
\ee

Setting  in (\ref{pent2a}) additionally $r=0$, $j=a^*$, $k=l$, $u_2=0$, $u_3=0$, $\beta_1=0$, $\gamma_1=\alpha_3$
we get

\be\label{pent3a}
\sum_{t_2}F_{0,l}\left[\begin{array}{cc}
a&b^*\\
a&b \end{array}\right]^{\gamma_1t_2}_{00}F_{b^*,0}\left[\begin{array}{cc}
a&a^*\\
l&l \end{array}\right]^{00}_{t_2\alpha_3}=F_{0,0}\left[\begin{array}{cc}
a&a^*\\
a&a \end{array}\right]^{00}_{00}\delta_{\gamma_1,\alpha_3}\equiv F_a\delta_{\gamma_1,\alpha_3}\, .
\ee

Setting in (\ref{pent1a}) $l=0$, $r=k^*$, $i^*=s$, $a=b$, $\gamma_1=0$, $t_2=0$,
$u_2=0$,  $u_3=t_2$ we receive

\be\label{pent4a}
\sum_{\beta_2}F_{p_2,i^*}\left[\begin{array}{cc}
j&k\\
p_1&a \end{array}\right]^{\beta_2u_3}_{\alpha_2\alpha_3}F_{p_1,0}\left[\begin{array}{cc}
i&i^*\\
a&a \end{array}\right]^{00}_{\alpha_1\beta_2}=\sum_{\beta_1}F_{p_1,k^*}\left[\begin{array}{cc}
i&j\\
a&p_2 \end{array}\right]^{\beta_1u_3}_{\alpha_1\alpha_2}
F_{p_2,0}\left[\begin{array}{cc}
k^*&k\\
a&a \end{array}\right]^{00}_{\beta_1\alpha_3}\, .
\ee
Setting in (\ref{pent4a}) $p_2=0$, $j=p_1$, $k=a^*$, $\alpha_2=0$, $\alpha_3=0$,
$\beta_1=0$, $\alpha_1=u_3$ we get
\be\label{pent44a}
\sum_{\beta_2}F_{0,i^*}\left[\begin{array}{cc}
j&a^*\\
j&a \end{array}\right]^{\beta_2u_3}_{00}F_{j,0}\left[\begin{array}{cc}
i&i^*\\
a&a \end{array}\right]^{00}_{\alpha_1\beta_2}=
F_a\delta_{\alpha_1,u_3}\, .
\ee
Eq.  (\ref{pent2a}) implies
\be\label{pent22a}
\sum_{\alpha_2}F_{0,p_1}\left[\begin{array}{cc}
j&p_2\\
j&p_2^* \end{array}\right]^{\gamma_1\alpha_2}_{00}F_{p_2,i^*}\left[\begin{array}{cc}
j&k\\
p_1&a \end{array}\right]^{\beta_2u_3}_{\alpha_2\alpha_3}=\sum_{\mu}F_{0,i^*}\left[\begin{array}{cc}
j&k\\
j&k^* \end{array}\right]^{\mu u_3}_{00}
F_{k^*,p_1}\left[\begin{array}{cc}
i^*&a\\
j&p_2^* \end{array}\right]^{\gamma_1\beta_2}_{\mu\alpha_3}\, .
\ee

Multiplying  (\ref{pent4a}) by $F_{0,p_1}\left[\begin{array}{cc}
j&p_2\\
j&p_2^* \end{array}\right]^{\gamma_1\alpha_2}_{00}$, summing by $\alpha_2$
and using (\ref{pent22a}) we derive

\bea\label{newthing}
\sum_{\mu,\beta_2}F_{0,i^*}\left[\begin{array}{cc}
j&k\\
j&k^* \end{array}\right]^{\mu u_3}_{00}
F_{k^*,p_1}\left[\begin{array}{cc}
i^*&a\\
j&p_2^* \end{array}\right]^{\gamma_1\beta_2}_{\mu\alpha_3}
F_{p_1,0}\left[\begin{array}{cc}
i&i^*\\
a&a \end{array}\right]^{00}_{\alpha_1\beta_2}=\\ \nonumber
\sum_{\beta_1,\alpha_2}F_{p_1,k^*}\left[\begin{array}{cc}
i&j\\
a&p_2 \end{array}\right]^{\beta_1u_3}_{\alpha_1\alpha_2}
F_{p_2,0}\left[\begin{array}{cc}
k^*&k\\
a&a \end{array}\right]^{00}_{\beta_1\alpha_3}F_{0,p_1}\left[\begin{array}{cc}
j&p_2\\
j&p_2^* \end{array}\right]^{\gamma_1\alpha_2}_{00}\, .
\eea

Eq. (\ref{pent3a}) and (\ref{pent44a}) imply

\be\label{pent3an}
\sum_{\alpha_1}F_{0,a}\left[\begin{array}{cc}
i&p_1\\
i&p_1^* \end{array}\right]^{\nu\alpha_1}_{00}F_{p_1,0}\left[\begin{array}{cc}
i&i^*\\
a&a \end{array}\right]^{00}_{\alpha_1\beta_2}= F_i\delta_{\nu,\beta_2}\, ,
\ee
\be\label{pent44an}
\sum_{\alpha_3}F_{0,k}\left[\begin{array}{cc}
p_2&a^*\\
p_2&a \end{array}\right]^{\alpha_3\rho}_{00}F_{p_2,0}\left[\begin{array}{cc}
k^*&k\\
a&a \end{array}\right]^{00}_{\beta_1\alpha_3}=
F_a\delta_{\beta_1,\rho}\, .
\ee

Multiplying (\ref{newthing}) by $F_{0,a}\left[\begin{array}{cc}
i&p_1\\
i&p_1^* \end{array}\right]^{\nu\alpha_1}_{00}$ and summing by $\alpha_1$ and then multiplying
by $F_{0,k}\left[\begin{array}{cc}
p_2&a^*\\
p_2&a \end{array}\right]^{\alpha_3\rho}_{00}$ and summing by $\alpha_3$, and using (\ref{pent3an}) and (\ref{pent44an}) we obtain

\bea\label{newthing2}
\sum_{\mu,\alpha_3}F_iF_{0,i^*}\left[\begin{array}{cc}
j&k\\
j&k^* \end{array}\right]^{\mu u_3}_{00}
F_{k^*,p_1}\left[\begin{array}{cc}
i^*&a\\
j&p_2^* \end{array}\right]^{\gamma_1\nu}_{\mu\alpha_3}
F_{0,k}\left[\begin{array}{cc}
p_2&a^*\\
p_2&a \end{array}\right]^{\alpha_3\rho}_{00}
=\\ \nonumber
\sum_{\alpha_1,\alpha_2}F_aF_{p_1,k^*}\left[\begin{array}{cc}
i&j\\
a&p_2 \end{array}\right]^{\rho u_3}_{\alpha_1\alpha_2}
F_{0,a}\left[\begin{array}{cc}
i&p_1\\
i&p_1^* \end{array}\right]^{\nu\alpha_1}_{00}
F_{0,p_1}\left[\begin{array}{cc}
j&p_2\\
j&p_2^* \end{array}\right]^{\gamma_1\alpha_2}_{00}\, .
\eea
Setting  in (\ref{newthing2}) $p_2=0$, $p_1=j$, $k=a^*$, $\gamma_1=0$, $\alpha_3=0$,
$\mu=\nu$, $\alpha_2=0$, $\rho=0$, $u_3=\alpha_1$ we get
\be\label{fio}
F_iF_{0,i^*}\left[\begin{array}{cc}
j&k\\
j&k^* \end{array}\right]^{\nu u_3}_{00}=F_{k^*}F_{0,k^*}\left[\begin{array}{cc}
i&j\\
i&j^* \end{array}\right]^{\nu u_3}_{00}\, .
\ee

Using (\ref{fio}) we obtain from (\ref{newthing2})
\bea\label{newthing23}
\sum_{\mu,\alpha_3}
F_{k^*}F_{0,k^*}\left[\begin{array}{cc}
i&j\\
i&j^* \end{array}\right]^{\mu u_3}_{00}
F_{k^*,p_1}\left[\begin{array}{cc}
i^*&a\\
j&p_2^* \end{array}\right]^{\gamma_1\nu}_{\mu\alpha_3}
F_{0,k}\left[\begin{array}{cc}
p_2&a^*\\
p_2&a \end{array}\right]^{\alpha_3\rho}_{00}
=\\ \nonumber
\sum_{\alpha_1,\alpha_2}
F_{p_1^*}F_{0,p_1^*}\left[\begin{array}{cc}
a^*&i\\
a^*&i^* \end{array}\right]^{\nu \alpha_1}_{00}
F_{p_1,k^*}\left[\begin{array}{cc}
i&j\\
a&p_2 \end{array}\right]^{\rho u_3}_{\alpha_1\alpha_2}
F_{0,p_1}\left[\begin{array}{cc}
j&p_2\\
j&p_2^* \end{array}\right]^{\gamma_1\alpha_2}_{00}\, .
\eea

Using   (\ref{fio}) one more time and the symmetries (\ref{symmmul}) we derive:

\bea\label{newthing234}
\sum_{\mu,\alpha_3}
F_{0,j}\left[\begin{array}{cc}
k&i\\
k&i^* \end{array}\right]^{ u_3\mu}_{00}
F_{k,p_1}\left[\begin{array}{cc}
i&a\\
j&p_2 \end{array}\right]^{\gamma_1\nu}_{\mu\alpha_3}
F_{0,k}\left[\begin{array}{cc}
p_2&a\\
p_2&a^* \end{array}\right]^{\rho\alpha_3}_{00}
=\\ \nonumber
\sum_{\alpha_1,\alpha_2}
F_{0,p_1}\left[\begin{array}{cc}
a&i\\
a&i^* \end{array}\right]^{\alpha_1\nu}_{00}
F_{p_1,k}\left[\begin{array}{cc}
i^*&j\\
a&p_2^* \end{array}\right]^{\rho u_3}_{\alpha_1\alpha_2}
F_{0,j}\left[\begin{array}{cc}
p_1&p_2\\
p_1&p_2^* \end{array}\right]^{\alpha_2\gamma_1}_{00}\, .
\eea

In the absence of the multiplicities Eq. (\ref{pent3a})  and   (\ref{fio})  take the forms:
\be\label{pent3aab}
F_{0,l}\left[\begin{array}{cc}
a&b^*\\
a&b \end{array}\right]F_{b^*,0}\left[\begin{array}{cc}
a&a^*\\
l&l \end{array}\right]=F_{0,0}\left[\begin{array}{cc}
a&a^*\\
a&a \end{array}\right]\equiv F_a\, ,
\ee
\be\label{fioab}
F_iF_{0,i^*}\left[\begin{array}{cc}
j&k\\
j&k^* \end{array}\right]=F_{k^*}F_{0,k^*}\left[\begin{array}{cc}
i&j\\
i&j^* \end{array}\right]\, .
\ee

Combining (\ref{pent3aab}),  (\ref{fioab})  we  get
\be\label{foifioa}
F_{0,i}\left[\begin{array}{cc}
j&k\\
j&k^* \end{array}\right]F_{i^*,0}\left[\begin{array}{cc}
k&k^*\\
j^*&j^* \end{array}\right]={F_jF_{k}\over F_i}\, .
\ee
In the absence of the multiplicities the symmetry properties  (\ref{symmmul}) take the form:
\be\label{symm}
F_{p,q}\left[\begin{array}{cc}
k&j\\
i&l \end{array}\right]=F_{p^*,q}\left[\begin{array}{cc}
j&k\\
l^*&i^* \end{array}\right]=F_{p,q^*}\left[\begin{array}{cc}
i^*&l\\
k^*&j \end{array}\right]=F_{p^*,q^*}\left[\begin{array}{cc}
l&i^*\\
j^*&k \end{array}\right]\, .
\ee

\section{Special functions}

{\bf The function} $\Gamma_b(x)$

The function $\Gamma_b(x)$ is a close relative of the double Gamma function studied in \cite{gam1,gam2}.
It can be defined by means of the integral representation 
\be
\log \Gamma_b(x)=\int_0^{\infty}{dt\over t}\left({e^{-xt}-e^{-Qt/2}\over (1-e^{-bt})(1-e^{-t/b})}-
{(Q-2x)^2\over 8e^t}-{Q-2x\over t}\right)\, .
\ee

Important properties of $\Gamma_b(x)$ are
\begin{enumerate}
\item 
Functional equation: $\Gamma_b(x+b)=\sqrt{2\pi}b^{bx-{1\over 2}}\Gamma^{-1}(bx)\Gamma_b(x)$.
\item
Analyticity:  $\Gamma_b(x)$ is meromorphic, poles: $x=-nb-mb^{-1}, n,m\in \mathbb{Z}^{\geq 0}$.
\item
Self-duality:  $\Gamma_b(x)= \Gamma_{1/b}(x)$.
\end{enumerate}

From the property 1 one can obtain the following relations:
\be
\Gamma_b(Q)=\sqrt{2\pi b}\Gamma_{1/b}\left({1\over b}\right)
\ee
\be
\Gamma_b(Q)=\sqrt{2\pi \over b}\Gamma_{1/b}(b)
\ee
\be\label{wgam}
W(x)=2^{-1/4}e^{3i\pi/2}{\Gamma_b(2x)\over \Gamma_b(2x-Q)}\lambda^{2x-Q\over 2b}\, ,
\ee
and the 
 behaviour of the  $\Gamma_b(x)$ near $x=0$:
\be
\Gamma_b(x)\sim {\Gamma_b(Q)\over 2\pi x}\, .
\ee

{\bf The function} $\Upsilon_b(x)$

The $\Upsilon_b$ may be defined in terms of $\Gamma_b$ as follows
\be
\Upsilon_b(x)={1\over  \Gamma_b(x) \Gamma_b(Q-x)}\, .
\ee

An integral representation convergent in the strip $0< {\rm Re}(x)<Q$ is
\be
\log \Upsilon_b(x)=\int_0^{\infty}{dt\over t}\left[\left({Q\over 2}-x\right)^2e^{-t}-{\sinh^2({Q\over 2}-x){t\over 2}\over
\sinh{bt\over 2}\sinh{t\over 2b}}\right]\, .
\ee
Important properties of $\Upsilon_b(x)$ are
\begin{enumerate}
\item 
Functional equation: $\Upsilon_b(x+b)=b^{1-2bx}{\Gamma(bx)\over \Gamma(1-bx)}\Upsilon_b(x)$.
\item
Analyticity:  $\Upsilon_b(x)$ is entire analytic, zeros: $x=-nb-mb^{-1}, n,m\in \mathbb{Z}^{\geq 0}$, $x=Q+nb+mb^{-1}, n,m\in \mathbb{Z}^{\geq 0}$\, .
\item
Self-duality:  $\Upsilon_b(x)= \Upsilon_{1/b}(x)$.

\end{enumerate}

These properties imply:
\be
{\Upsilon_b(2x)\over \Upsilon_b(2x-Q)}=S(x)\lambda^{{2x-Q\over b}}
\ee
and
\be
\Upsilon_b(x)\sim x\Upsilon_b(b), 
\ee
when $x\rightarrow 0$.

{\bf The function} $S_b(x)$

The function $S_b(x)$ may be defined in terms of $\Gamma_b(x)$ as follows 
\be
S_b(x)={\Gamma_b(x)\over \Gamma_b(Q-x)}\, .
\ee

An integral that represents $\log S_b(x)$ is
\be
\log S_b(x)=\int_0^{\infty} {dt\over t}\left({\sinh t(Q-2x)\over 2\sinh bt\sinh b^{-1}t}-{Q-2x\over 2t}\right)\, .
\ee
The most important properties are
\begin{enumerate}
\item 
Functional equation:   $S_b(x+b)=2\sin \pi bx S_b(x)$\, .
\item
Analiticity: $S_b(x)$ is meromorphic, poles: $x=-(nb+mb^{-1})$, $n,m\in {\mathbb Z}^{\geq 0}$\, , zeros $x=Q+(nb+mb^{-1})$,
$n,m\in {\mathbb Z}^{\geq 0}$\, .
\item
Self-duality: $S_b(x)=S_{1/b}(x)$\, .
\item
Inversion relation: $S_b(x)S_b(Q-x)=1$\, .
\end{enumerate}

These properties imply:
\be
{S_b(2x)\over S_b(2x-Q)}=-\sqrt{2}W(x)W(Q-x)
\ee


\newpage


\begin{thebibliography}{99}
\bibitem{Alday:2009aq}
  L.~F.~Alday, D.~Gaiotto and Y.~Tachikawa,
  ``Liouville Correlation Functions from Four-dimensional Gauge Theories,''
  Lett.\ Math.\ Phys.\  {\bf 91} (2010) 167
  [arXiv:0906.3219 [hep-th]].
\bibitem{Alday:2009fs}
  L.~F.~Alday, D.~Gaiotto, S.~Gukov, Y.~Tachikawa and H.~Verlinde,
  ``Loop and surface operators in N=2 gauge theory and Liouville modular
  geometry,''
  JHEP {\bf 1001} (2010) 113
  [arXiv:0909.0945 [hep-th]].
\bibitem{gam1}
E.~ W.~ Barnes, ``Theory of the double gamma function", Phil.\  Trans.\ Roy.\  Soc {\bf A196} (1901) 265-388

\bibitem{Behrend:1999bn}
  R.~E.~Behrend, P.~A.~Pearce, V.~B.~Petkova and J.~B.~Zuber,
  ``Boundary conditions in rational conformal field theories,''
  Nucl.\ Phys.\  B {\bf 570} (2000) 525
  [Nucl.\ Phys.\  B {\bf 579} (2000) 707]
  [arXiv:hep-th/9908036].
\bibitem{Belavin:1984vu}
  A.~A.~Belavin, A.~M.~Polyakov and A.~B.~Zamolodchikov,
  ``Infinite conformal symmetry in two-dimensional quantum field theory,''
  Nucl.\ Phys.\  B {\bf 241} (1984) 333.




\bibitem{Braaten:1982yn}
  E.~Braaten, T.~Curtright and C.~B.~Thorn,
  ``An Exact Operator Solution Of The Quantum Liouville Field Theory,''
  Annals Phys.\  {\bf 147} (1983) 365.
\bibitem{Braaten:1983np}
  E.~Braaten, T.~Curtright, G.~Ghandour and C.~B.~Thorn,
  ``Nonperturbative Weak Coupling Analysis Of The Quantum Liouville Field
  Theory,''
  Annals Phys.\  {\bf 153} (1984) 147.
\bibitem{Curtright:1982gt}
  T.~L.~Curtright and C.~B.~Thorn,
  ``Conformally Invariant Quantization Of The Liouville Theory,''
  Phys.\ Rev.\ Lett.\  {\bf 48} (1982) 1309
  [Erratum-ibid.\  {\bf 48} (1982) 1768].
\bibitem{Cardy:1989ir}
  J.~L.~Cardy,
  ``Boundary Conditions, Fusion Rules and the Verlinde Formula,''
  Nucl.\ Phys.\  B {\bf 324} (1989) 581.

\bibitem{Dorn:1994xn}
  H.~Dorn and H.~J.~Otto,
  ``Two and three point functions in Liouville theory,''
  Nucl.\ Phys.\ B {\bf 429} (1994) 375
  [hep-th/9403141].
\bibitem{Dotsenko:1984ad}
  V.~S.~Dotsenko and V.~A.~Fateev,
  ``Four Point Correlation Functions And The Operator Algebra In The
  Two-Dimensional Conformal Invariant Theories With The Central Charge $C < 1$,''
  Nucl.\ Phys.\  B {\bf 251} (1985) 691.
\bibitem{Drukker:2010jp}
  N.~Drukker, D.~Gaiotto and J.~Gomis,
  ``The Virtue of Defects in 4D Gauge Theories and 2D CFTs,''
  JHEP {\bf 1106} (2011) 025
  [arXiv:1003.1112 [hep-th]].
\bibitem{Drukker:2009id}
  N.~Drukker, J.~Gomis, T.~Okuda and J.~Teschner,
  ``Gauge Theory Loop Operators and Liouville Theory,''
  JHEP {\bf 1002} (2010) 057
  [arXiv:0909.1105 [hep-th]].


\bibitem{Fateev:2007ab}
  V.~A.~Fateev and A.~V.~Litvinov,
  ``Correlation functions in conformal Toda field theory. I,''
  JHEP {\bf 0711} (2007) 002
  [arXiv:0709.3806 [hep-th]].
\bibitem{Fateev:2000ik}
  V.~Fateev, A.~B.~Zamolodchikov and A.~B.~Zamolodchikov,
  ``Boundary Liouville field theory. I: Boundary state and boundary  two-point
  arXiv:hep-th/0001012.
\bibitem{Fateev:2001mj}
  V.~A.~Fateev,
  ``Normalization factors, reflection amplitudes and integrable systems,''
  arXiv:hep-th/0103014.
\bibitem{Fateev:2010za}
  V.~Fateev and S.~Ribault,
  ``Conformal Toda theory with a boundary,''
  JHEP {\bf 1012} (2010) 089
  [arXiv:1007.1293 [hep-th]].
\bibitem{Felder:1989hq}
  G.~Felder, J.~Frohlich and G.~Keller,
  ``On the structure of unitary conformal field theory. 2. Representation
  theoretic approach,''
  Commun.\ Math.\ Phys.\  {\bf 130} (1990) 1.
\bibitem{Felder:1989wv}
  G.~Felder, J.~Frohlich and G.~Keller,
  ``Braid matrices and structure constant for minimal conformal models,''
  Commun.\ Math.\ Phys.\  {\bf 124} (1989) 647.
\bibitem{Fuchs:2002cm}
  J.~Fuchs, I.~Runkel and C.~Schweigert,
  ``TFT construction of RCFT correlators. I: Partition functions,''
  Nucl.\ Phys.\  B {\bf 646} (2002) 353
  [arXiv:hep-th/0204148].
\bibitem{Fuchs:2004xi}
  J.~Fuchs, I.~Runkel and C.~Schweigert,
  ``TFT construction of RCFT correlators IV: Structure constants and
  Nucl.\ Phys.\  B {\bf 715} (2005) 539
  [arXiv:hep-th/0412290].
\bibitem{Gomis:2010kv}
  J.~Gomis and B.~Le Floch,
  ``'t Hooft Operators in Gauge Theory from Toda CFT,''
  arXiv:1008.4139 [hep-th].
\bibitem{Harlow:2011ny}
  D.~Harlow, J.~Maltz and E.~Witten,
  ``Analytic Continuation of Liouville Theory,''  JHEP {\bf 1112} (2011) 071  [arXiv:1108.4417 [hep-th]].  


\bibitem{Moore:1988ss}
  G.~W.~Moore and N.~Seiberg,
  ``Naturality in Conformal Field Theory,''
  Nucl.\ Phys.\  B {\bf 313} (1989) 16.
\bibitem{Moore:1989vd}
  G.~W.~Moore and N.~Seiberg,
  ``Lectures on RCFT,''  Published in Trieste Superstrings 1989:1-129.  Also in Banff NATO ASI 1989:263-362.




\bibitem{Moore:1988qv}
  G.~W.~Moore and N.~Seiberg,
  ``Classical and Quantum Conformal Field Theory,''
  Commun.\ Math.\ Phys.\  {\bf 123} (1989) 177.

\bibitem{Passerini:2010pr}
  F.~Passerini,
  ``Gauge Theory Wilson Loops and Conformal Toda Field Theory,''
  JHEP {\bf 1003} (2010) 125
  [arXiv:1003.1151 [hep-th]].
\bibitem{Petkova:2000ip}
  V.~B.~Petkova and J.~B.~Zuber,
  ``Generalized twisted partition functions,''
  Phys.\ Lett.\  B {\bf 504} (2001) 157
  [arXiv:hep-th/0011021].


\bibitem{Petkova:2009pe}
  V.~B.~Petkova,
  ``On the crossing relation in the presence of defects,''
  JHEP {\bf 1004} (2010) 061
  [arXiv:0912.5535 [hep-th]].


\bibitem{Ponsot:1999uf}
  B.~Ponsot and J.~Teschner,
  ``Liouville bootstrap via harmonic analysis on a noncompact quantum group,''
  arXiv:hep-th/9911110.


\bibitem{Ponsot:2001ng}
  B.~Ponsot and J.~Teschner,
  ``Boundary Liouville field theory: Boundary three point function,''
  Nucl.\ Phys.\  B {\bf 622} (2002) 309
  [arXiv:hep-th/0110244].
\bibitem{Recknagel:2002qq}
  A.~Recknagel,
  ``Permutation branes,''
  JHEP {\bf 0304} (2003) 041
  [arXiv:hep-th/0208119].


\bibitem{Runkel:1998pm}
  I.~Runkel,
  ``Boundary structure constants for the A series Virasoro minimal models,''
  Nucl.\ Phys.\  B {\bf 549} (1999) 563
  [arXiv:hep-th/9811178].
\bibitem{Sarkissian:2009aa}
  G.~Sarkissian, ``Defects and Permutation branes in the Liouville field theory,''  Nucl.\ Phys.\ B {\bf 821} (2009) 607-625
  arXiv:0903.4422 [hep-th].
\bibitem{gam2}
T.~ Shintani, ``On a Kronecker limit formula for real quadratic fields", J.\  Fac.\ Sci.\ Univ.\ Tokyo Sect. 1A Math.
{\bf 24} (1977) 167-199

\bibitem{Schomerus:2005aq}
  V.~Schomerus,
  ``Non-compact string backgrounds and non-rational CFT,''
  Phys.\ Rept.\  {\bf 431} (2006) 39
  [hep-th/0509155].
\bibitem{Teschner:2001rv}
  J.~Teschner,
  ``Liouville theory revisited,''
  Class.\ Quant.\ Grav.\  {\bf 18} (2001) R153
  [arXiv:hep-th/0104158].
\bibitem{Teschner:2008qh}
  J.~Teschner,
  ``Nonrational conformal field theory,''
  arXiv:0803.0919 [hep-th].
\bibitem{Teschner:1995yf}
  J.~Teschner,
  ``On the Liouville three point function,''
  Phys.\ Lett.\  B {\bf 363} (1995) 65
  [arXiv:hep-th/9507109].
\bibitem{Zamolodchikov:1995aa}
  A.~B.~Zamolodchikov and A.~B.~Zamolodchikov,
  ``Structure constants and conformal bootstrap in Liouville field theory,''
  Nucl.\ Phys.\ B {\bf 477} (1996) 577
  [hep-th/9506136].
\bibitem{Zamolodchikov:2001ah}
  A.~B.~Zamolodchikov and A.~B.~Zamolodchikov,
  ``Liouville field theory on a pseudosphere,''
  arXiv:hep-th/0101152.


\end{thebibliography}
\end{document}